\documentclass{article}

\usepackage{arxiv}

\usepackage{cite}
\usepackage{amsmath,amssymb,amsfonts}
\usepackage{algorithmic}
\usepackage{graphicx}
\usepackage{textcomp}
\usepackage{xcolor}
\usepackage[T1]{fontenc}
\usepackage[latin9]{inputenc}
\usepackage{array}
\usepackage{float}
\usepackage{nomencl}
\usepackage{multirow}
\usepackage{dirtytalk}
\usepackage{tabto}
\usepackage{comment}
\usepackage{subcaption}
\usepackage{adjustbox}
\usepackage{amsmath}
\usepackage{adjustbox}
\usepackage{makecell}
\usepackage{longtable}
\usepackage{mathtools}
\usepackage{array,ragged2e}
\usepackage{booktabs}
\usepackage{hyperref}

\title{Automated CVE Analysis for Threat Prioritization and Impact Prediction}

\author{ {Ehsan Aghaei and Ehab Al-Shaer} \\
	Department of Computer Science\\
	Carnegie Mellon University\\
	Pittsburgh, PA \\
	\texttt{\{eaghaei,ealshaer\}@andrew.cmu.edu} \\
 	\And
	{Waseem Shadid and Xi Niu} \\
	Department of Computing and Information Systems\\
	University of North Carolina - Charlotte\\
	Charlotte, NC \\
	\texttt{\{wshadid, xniu2\}@charlotte.edu} \\
}
\date{}

\begin{document}

\maketitle

\hypersetup{
pdftitle={A template for the arxiv style},
pdfsubject={q-bio.NC, q-bio.QM},
pdfauthor={David S.~Hippocampus, Elias D.~Striatum},
pdfkeywords={First keyword, Second keyword, More},
}

\begin{abstract}
The Common Vulnerabilities and Exposures (CVE) are pivotal information for proactive cybersecurity measures, including service patching, security hardening, and more. However, CVEs typically offer low-level, product-oriented descriptions of publicly disclosed cybersecurity vulnerabilities, often lacking the essential attack semantic information required for comprehensive weakness characterization and threat impact estimation. This critical insight is essential for CVE prioritization and the identification of potential countermeasures, particularly when dealing with a large number of CVEs.

Current industry practices involve manual evaluation of CVEs to assess their attack severities using the Common Vulnerability Scoring System (CVSS) and mapping them to Common Weakness Enumeration (CWE) for potential mitigation identification. Unfortunately, this manual analysis presents a major bottleneck in the vulnerability analysis process, leading to slowdowns in proactive cybersecurity efforts and the potential for inaccuracies due to human errors.

In this research, we introduce our novel predictive model and tool (called CVEDrill) which revolutionizes CVE analysis and threat prioritization. CVEDrill accurately estimates the CVSS vector for precise threat mitigation and priority ranking and seamlessly automates the classification of CVEs into the appropriate CWE hierarchy classes. By harnessing CVEDrill, organizations can now implement cybersecurity countermeasure mitigation with unparalleled accuracy and timeliness, surpassing in this domain the capabilities of state-of-the-art tools like ChaptGPT.\\
\end{abstract}

\section{Introduction}
The CVE (Common Vulnerabilities and Exposures) system plays a crucial role in the cybersecurity community by facilitating the disclosure and sharing of software and hardware vulnerabilities with the potential for cyberattacks~\cite{(2)CveMITRE2018}. CVEs provide concise descriptions of vulnerabilities, aiding security professionals in understanding their nature. However, these descriptions often lack the necessary attack semantic information required for comprehensive weakness characterization and threat impact estimation, making prioritization and countermeasure identification challenging, especially when dealing with a large number of CVEs.

Given the significant number of unpatched vulnerabilities that can persist for extended periods, understanding their priority based on impact and identifying inherent exploitable weaknesses and potential countermeasures are critical goals for proactive defense. Consequently, deep analysis of CVEs is essential to grasp vulnerability characteristics and implications effectively.

The cybersecurity community has dedicated substantial efforts over the years to establish standards and methodologies that enhance the comprehension and characterization of CVEs. Notably, the Common Vulnerability Scoring System (CVSS) emerged as a standardized framework for evaluating vulnerability severity and exploitability, enabling effective threat prioritization. CVSS, an open industry standard, assigns numerical scores to CVEs, which can be translated into qualitative representations like low, medium, high, and critical, aiding organizations in prioritizing response and mitigation measures.

The CVSS framework encompasses three metric groups: Base Metrics, which evaluate vulnerability characteristics like attack vector and complexity independently of the environment; Temporal Metrics, which account for factors like exploit code availability, remediation level, and report confidence that may vary over time; and Environmental Metrics, which consider specific environmental factors such as system impact and asset importance.

Additionally, the cybersecurity community introduced a higher-level abstraction known as Common Weakness Enumeration (CWE) to represent CVEs~\cite{(3)CweMITRE2018}. CWE serves as a community-developed catalog of prevalent software and hardware weaknesses, describing the root causes of vulnerabilities. It operates as a hierarchically-designed dictionary of software flaws, assisting in understanding their potential impacts and recommending practices and mitigations to address specific weaknesses and vulnerabilities.

Present industry practices necessitate manual handling of both CVSS assessments and CVE classification into specific CWE classes. Regrettably, this manual approach creates a significant bottleneck in the vulnerability analysis process, resulting in delays in proactive cybersecurity endeavors and the possibility of inaccuracies caused by human errors. Furthermore, the manual assessment and analysis of CVEs often proves labor-intensive and inconsistent. The absence of security automation in this context forces security analysts to manually analyze, investigate, and evaluate the appropriate mitigation measures for a substantial number of vulnerabilities on a daily basis.

This paper introduces our research, focused on developing techniques and tools to automate the predicting the core metrics composing the CVSS scores, and classification to CWE for any new and unexplored CVE solely by analyzing its unstructured text description. Our CVSS predictor and CVE classifier implemented in a tool called {\em CVEDrill} that leverages domain-specific language models for cybersecurity, enabling computers to interpret qualitative input and convert it into quantitative representations.

Language models, a form of artificial intelligence, are designed to comprehend and generate human language. Trained on vast text data, they learn to predict word probabilities given context. In our research, language models prove invaluable for understanding cybersecurity sentences and phrases, bridging the semantic gap in unstructured text and inferring concept correlations that may elude human analysis. This capability empowers powerful text prediction and classification, leading to automated cybersecurity decision-making in a timely and cost-effective manner.

Our novel predictive model (CVEDrill) revolutionizes CVE analysis and threat prioritization. CVEDrill accurately estimates the CVSS vector and automates the classification of CVEs into appropriate CWE classes, providing precise threat priority ranking, and mitigation. By harnessing CVEDrill, organizations can now implement cybersecurity countermeasure mitigation with unparalleled accuracy and efficiency, surpassing even state-of-the-art tools like ChatGPT.

Unlike ChatGPT, CVEDrill continuously learns and refines its capabilities, ensuring high accuracy in vulnerability assessments through its automated CVSS predictor. Additionally, ChatGPT may prove resource-intensive and economically expensive. Therefore, our approach to developing lightweight, economical domain-specific language models offers a more practical and effective solution for cybersecurity tasks.


This paper briefly discusses our domain-specific language model for cybersecurity in Section \ref{sec:SecureBERT: A Domain-Specific Language Model}. Then, it describes our methodologies to automatically predict the CVSS base metrics for CVEs, as well as classify CVEs to CWEs in Section \ref{sec:CVSS Prediction} and \ref{sec: Automated CVE to CWE Classification}, respectively. Finally, we review the existing relaed work in Section \ref{sec:Related Works} and discuss their advantages and limitations.

\section{SecureBERT: A Domain-Specific Language Model}\label{sec:SecureBERT: A Domain-Specific Language Model}
Language models are developed to represent the most statistically important features of the distribution of word sequences in a natural language, typically enabling probabilistic predictions of the following word given the preceding ones.
Typically, traditional techniques require creating an n$^{th}$ order Markov assumption and estimating n-gram probabilities via counting and subsequently smoothing. Although these statistical models are straightforward to train, the probability of rare n-grams can be underestimated due to data sparsity, even when smoothing techniques are used.
Neural Language Models (NLM) employ word embedding vector representations to encode words as inputs to neural networks to cope with the sparsity of n-gram features.

The capability of machines to interpret qualitative data and transform it into quantitative information, which the machines may subsequently employ for any underlying tasks, makes language modeling vital in mainstream NLP applications.
There are several well-known and well-performing language models, such as ELMO~\cite{peters2018deep}, GPT~\cite{radford2018improving}, and BERT~\cite{devlin2018bert}, trained on general English corpora and used for a variety of NLP tasks such as machine translation, named entity recognition, text classification, and semantic analysis. 
However, certain domains, such as cybersecurity, are indeed highly sensitive, dealing with the processing of critical data and any error in the process may expose the entire infrastructure to cyber threats, and therefore, automated processing of cybersecurity text requires a robust and reliable framework trained on domain-specific corpora.  Cybersecurity terms are either uncommon in general English (e.g., \textit{ransomware, API, OAuth, exfilterate}, and \textit{keylogger}) or have a different meaning (homographs) in other domains (such as \textit{run, honeypot, patch, handshake}, and \textit{worm}).
The existing structural and semantic gap across different domains affects text processing and demonstrates that the standard English language model may not be able to accommodate the vocabulary of cybersecurity texts, resulting in a shallow grasp of the implications of cybersecurity.

BERT (Bidirectional Encoder Representations from Transformers) \cite{devlin2018bert} is a transformer-based neural network technique for natural language processing pre-training. BERT can train language models based on the entire set of words in a sentence or query (bidirectional training) rather than the traditional training on the ordered sequence of words (left-to-right or combined left-to-right and right-to-left). BERT allows the language model to learn word context based on surrounding words rather than just the word that immediately precedes or follows it.
SecureBERT~\cite{aghaei2022securebert} is a cybersecurity domain-specific language model built on top of the RoBERTa\cite{liu2019roberta}, a derivative of BERT, which has been retrained on a large number of cybersecurity documents. RoBERTa noticeably improved on the MLM and accordingly the overall performance in all standard fine-tuning tasks compared to BERT. SecureBERT,  which employs a cybersecurity-based tokenizer, has demonstrated promising performance in various cybersecurity-related tasks including masked language modeling (MLM), named entity recognition (NER), and sentiment analysis. Therefore, we utilized this pre-trained language model for our prediction tasks.
Pre-trained SecureBERT takes the tokenized text as input and passes it through the transformers stack and returns the embedding representation of each token in addition to an extra vector called [CLS]. In contrast to the hidden state vector associated with any other token, the hidden state associated with [CLS] is an aggregate representation of the entire input used for classification tasks.
Depending on the nature of the fine-tuning task, a predictive layer can be set on top of the SecureBERT taking [CLS] or any other set of pre-trained model outputs for further process.
\begin{figure}
 \center
  \includegraphics[width=6.5cm]{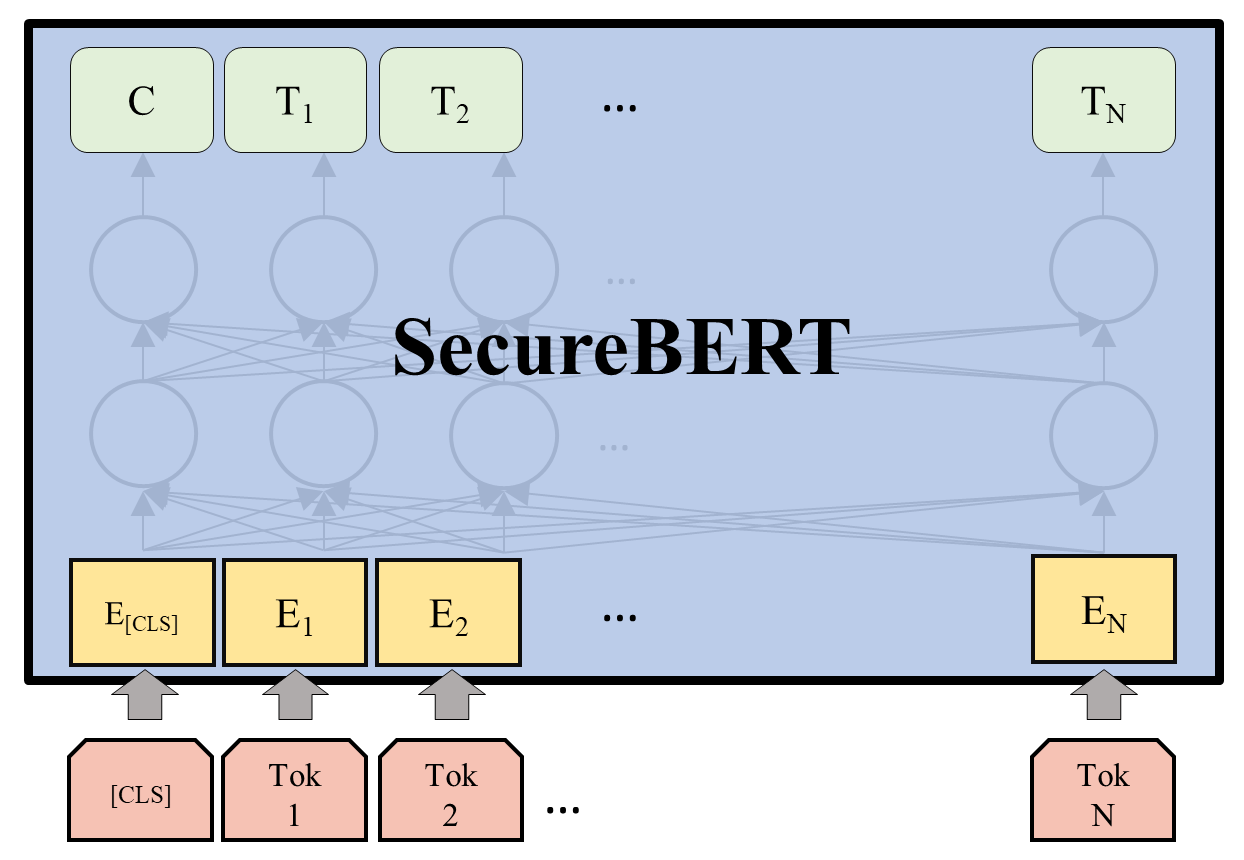}
  \caption{The general architecture of SecureBERT which takes the tokenized input and returns the corresponding set of vector representations.}
  \label{fig: cvss_model}
\end{figure}

\section{CVSS Prediction}\label{sec:CVSS Prediction}
When it comes to documenting the severity of software vulnerabilities, CVSS is the standard protocol, which has been issued in two versions, with CVSS V2, and CVSS V3 (and CVSS V3.1) \footnote{https://www.first.org/cvss/} both being extensively utilized in the computer security domain.  
Both CVSS versions are structured similarly, with a series of multiple-choice questions to the vendors concerning the vulnerability and classified into three groups: base, temporal, and environmental. The base group defines the inherent properties of the vulnerability, the temporal group defines how those properties vary over time, and the environmental group illustrates the severity of a vulnerability in the context of a particular organization. In this study, our focus will be on CVSS V3, as it is more recent and has seen wider adoption compared to the previous version

\begin{table*}
\scriptsize
\centering
 \begin{tabular}{|l | l | l |} 
  \hline
 \multicolumn{3}{c}{\textbf{CVSS V3}} \\
  \hline
  \hline
   \multicolumn{1}{|c|}{\textbf{Base Vector}} & \multicolumn{1}{|c|}{\textbf{Temporal Vector}} & \multicolumn{1}{|c|}{\textbf{Environmental Vector}}\\
   \hline
  Attack Vector (AV) &  &\\
  Attack Complexity (AC) &  &Modified Base Metrics (M*)\\
  Privileges Required (PR) & Exploit Code Maturity (E) & Confidentiality \\
  User Interaction (UI) & Remediation Level (RL)& Requirement (CR)\\
  Confidentiality (C) & Report Confidence (RC)   &Integrity Requirement (IR) \\
  Availability (A) & & Availability Requirement (AR)\\
  Integrity (I) & & \\
  Scope (S) &  &\\
   \hline
\end{tabular}\hfil
     \caption{\label{tab: CVSSV3} The resources collected for cybersecurity textual data}
\end{table*}

Table \ref{tab: CVSSV3} shows the different fields required by CVSS V3. When all responses about a vulnerability have been completed, they are merged using a common syntax to form a CVSS Vector. Each CVSS version has a non-linear severity formula that accepts all data as input and delivers a severity value between 0.0 and 10.0.
Since this equation is not linear, slight changes in the vector can result in significant variances in the severity score. CVSS V3 and V3.1 have the same fields and differ solely in the severity algorithm, which affects a small number of vulnerabilities. In the rest of our work, we use the term "CVSS V3" to describe both the CVSS V3 and V3.1 specifications. In addition, since  Temporal and Environmental vectors are derivable from Base Vector, we only process the Base vector, which consists of eight different metrics including \textit{Attack Vector (AV)}, \textit{Attack Complexity (AC)}, \textit{Privileges Required (PR)}, \textit{User Interaction (UI)}, \textit{Confidentiality (C)}, \textit{Availability (A)}, \textit{Integrity (I)},  and \textit{Scope (S)}. Each metric also can get different values which are demonstrated in Table \ref{tab: CVSSV3_base}. Our target in this work is to automatically predict the value of each of the eight metrics for a given vulnerability using the corresponding CVE ID description, utilizing a pre-trained SecureBERT language model.

\begin{table*}
\footnotesize
\centering
 \begin{tabular}{|l | l |} 
\hline
 \multicolumn{2}{c}{\textbf{CVSS V3 Base Metric}} \\
  \hline
  \hline
   \multicolumn{1}{|c|}{\textbf{Metric}} & \multicolumn{1}{|c|}{\textbf{Values (classes)}} \\
   \hline
  Attack Vector (AV) & Network (N) | Adjacent (A) | Local (L) | Physical (P) \\
  \hline
  Attack Complexity (AC) & Low (L) | High (H) \\\hline
  Privileges Required (PR) & None (N) | Low (L) | High (H) \\\hline
  User Interaction (UI) & None (N) | Required (R)\\\hline
  Confidentiality (C) & None (N) | Low (L) | High (H) \\\hline
  Availability (A) & None (N) | Low (L) | High (H)\\\hline
  Integrity (I) &  None (N) | Low (L) | High (H)\\\hline
  Scope (S) & Unchanged (U) | Changed (C) \\
   \hline
\end{tabular}\hfil
     \caption{\label{tab: CVSSV3_base} {Shows the CVSS V3 Base metrics and the potential values.}}
\end{table*}

Currently, the CVSS is calculated through manual engineering effort that is expensive, inefficient, inconsistent, and problematic. The NIST's security experts take days or even longer to analyze CVEs and measure their severities such that many CVEs may remain indeterminate. This controversial cycle simply means cybersecurity analysts cannot rely on the availability of perpetual severity metrics for every CVE. Thus, they are limited to available CVE elements such as description to focus. 
By automating the process of predicting CVSS scores, organizations can effectively and accurately evaluate vulnerability severity, prioritize remediation efforts, and make well-informed decisions regarding allocating security resources. This automation empowers organizations to promptly respond to emerging threats, improve their vulnerability management procedures, and ultimately enhance their security.

\subsection{Challenges}
Automating CVSS vector prediction is a challenging task. The existing works are limited or built upon imprecise assumptions. Predicting the CVSS score requires a deep analysis of each vector element. Therefore, a direct approach to obtaining the final severity score would be problematic since any slight mistake may significantly compromise the output. Alternatively, breaking down the CVSS into smaller components and addressing the issue incrementally would yield comprehensive insights into the vulnerability. This approach gains significance as it's been established that vulnerability explanations frequently lack complete details regarding threat attributes. Consequently, possessing partial information about the threat proves advantageous when the automated model struggles to deliver precise values for every metric within the CVSS vector.

 employing conventional techniques for information extraction and machine learning algorithms might not offer the most optimal strategy for addressing this challenge. These approaches primarily rely on statistical analysis and word frequency, treating the provided text as a bag-of-word, which fails to comprehensively grasp the context and extract meaningful connections within the text, resulting in the omission of crucial details and, consequently, an inadequately resilient predictive model.

In addition, the current labeled CVSS dataset furnished by NVD displays a notable imbalance data problem, indicating a substantial variation in the occurrence frequency of values within individual metrics.
Previous studies have not effectively tackled the challenge posed by imbalanced data and have not conducted a comprehensive performance assessment of this concern within their proposed methodologies.

\begin{figure*}
\centerline{\includegraphics [width=13cm]{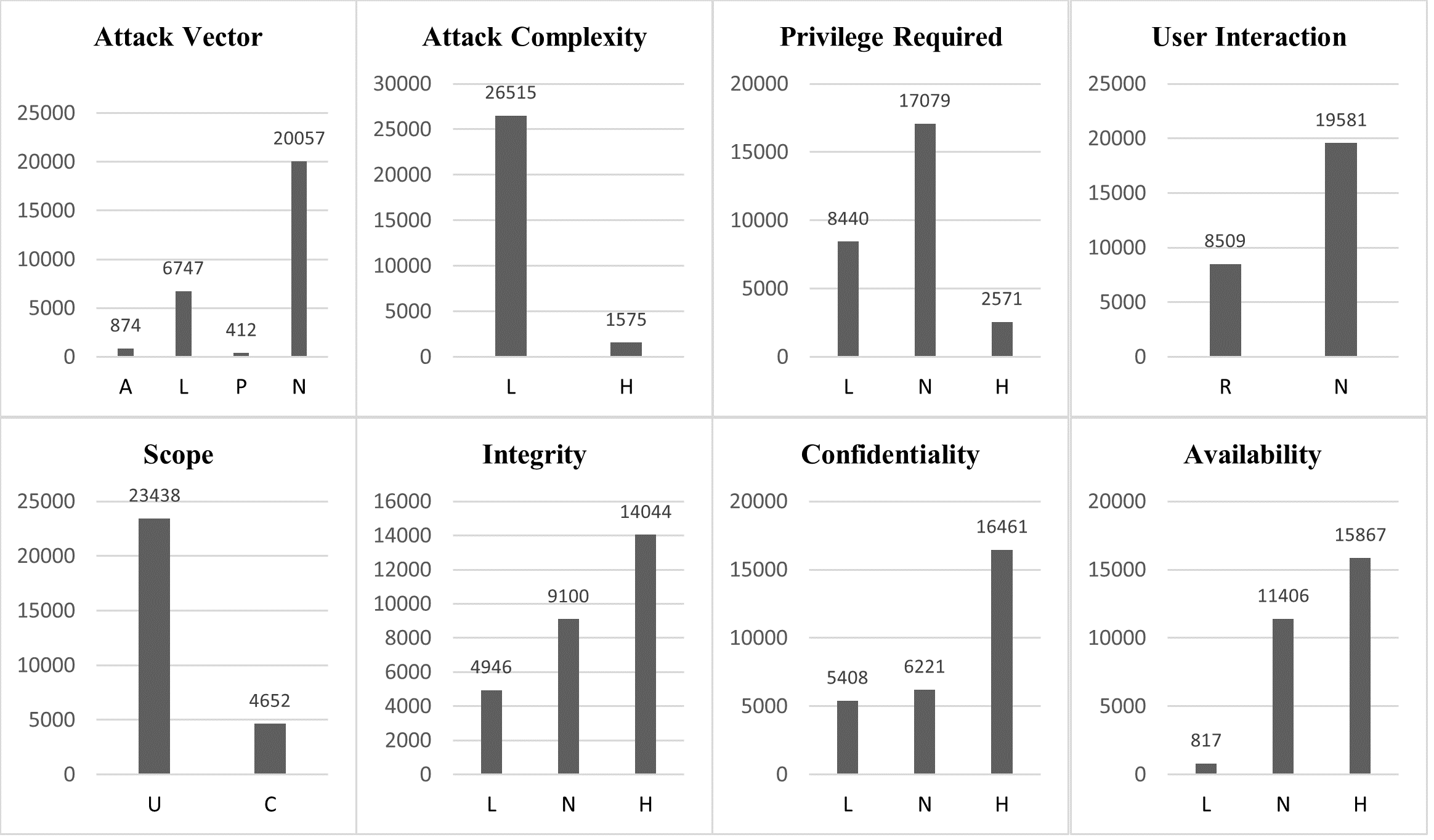}}
\caption {Numbers of available labeled records in each value for each CVSS metric. This shows the imbalance data problem in the CVSS dataset}
\label{fig: metric_dist}
\end{figure*}

Table \ref{fig: metric_dist} shows the value distribution within each metric. 

\subsection{Methodology}
In this section, we outline our approach to predicting the CVSS vector. Given the presence of eight metrics, we develop eight separate models. These models utilize the vulnerability's CVE description as input and generate the projected value for the corresponding CVSS metric as output. A general overview of the proposed CVSS analysis pipeline is depicted in Fig. \ref{fig: cvss_model}.
We incorporated the pre-trained SecureBERT model \cite{aghaei2022securebert} as the foundational framework, and introduced an additional TF-IDF module to address the issue of imbalanced data. In this architectural design, the CVE description is tokenized using SecureBERT's BPE tokenizer \cite{bostrom2020byte}, then processed through the transformer layers. Subsequently, a pooling layer is applied to transform the vector representation of the [CLS] token, which is then merged with a distinct vector (tensor) representing a customized TF-IDF vector derived from the same CVE description. This combined output vector has a dimension equal to the total possible values for the specific metric of interest and is ultimately implemented as a feed-forward neural network through an output-dense layer.
\begin{figure}
 \center
  \includegraphics[width=8.5cm]{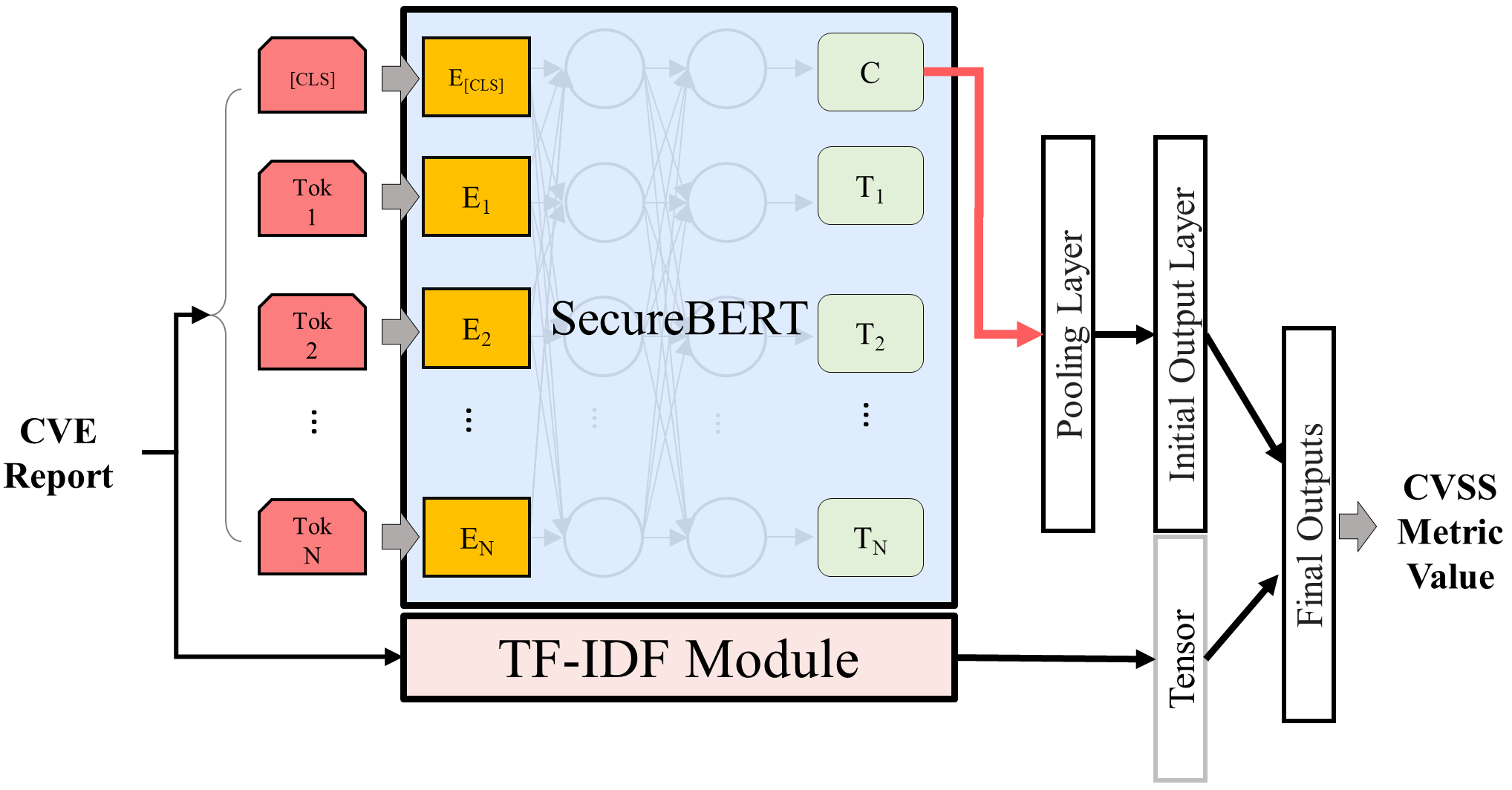}
  \caption{CVSS metric value prediction model design. Each CVSS metric has a separate model that is trained independently.}
  \label{fig: cvss_model}
\end{figure}

As previously noted, the CVSS data exhibits a significant imbalance. To mitigate the influence of this problem during training, we employ a personalized TF-IDF~\cite{ramos2003using} module. This module generates vectors by considering word frequencies, thereby representing input data associated with the minority classes (values) in the models dedicated to each metric. In essence, our research reveals that certain words, termed "signature words," recurrently appear within the minority classes when each potential value within a metric is regarded as a distinct class. For example, in CVEs whose the value of \textit{Attack Vector (AV)} metric is \textit{Adjacent Network (A)}, words like "adjacent," "Bluetooth," "pairing," and "dongle" are frequently appeared in the description. In another example, when the value of \textit{Attack Complexity (AC)} is \textit{High (H)}, "man-in-the-middle," "memory-cache," "padding," and "prolog" are frequently observed. 
On the other hand, keyword matching for predicting the value would not be a practical approach as these words do not uniquely appear in one specific class. In addition, when the training data is not large enough, even a powerful language model will not be able to learn the context and semantic relationships sufficiently.

TF-IDF, short for term frequency-inverse document frequency, is a statistical method that evaluates the importance of a term within a collection of documents. It assigns numerical values to words based on their significance in a document, aiding in the recognition of influential words that are infrequently found in other documents. These distinctive words, known as "signature words," hold significant meaning within specific categories or contexts.
Our utilization of TF-IDF scores aims to enhance the model's ability to identify minority classes during training. This involves extracting signature words and converting them into vectors for integration with SecureBERT's output. By focusing on crucial yet less common signature keywords during training, we expect the model to better distinguish minority classes within metrics. Each value within a metric is regarded as a distinct class.

\begin{figure*}
 \center
  \includegraphics[width=0.8\textwidth]{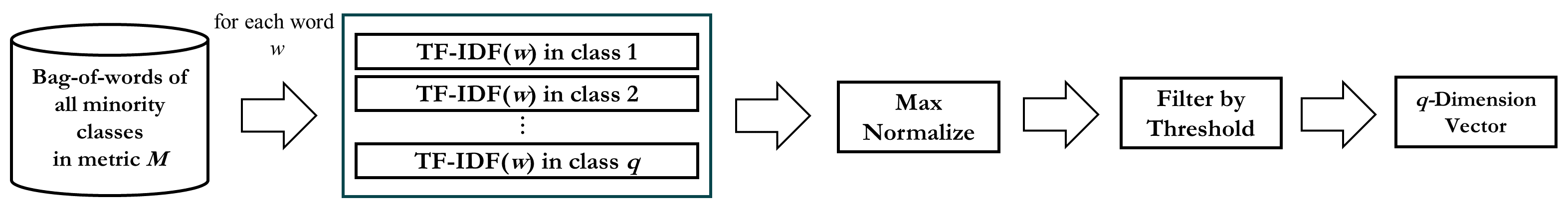}
  \caption{Shows the different steps in generating customized TF-IDF vectors to represent CVE descriptions.}
  \label{fig: TF-IDF_schema}
\end{figure*}

In this context, we first conduct the standard text cleaning steps on the training dataset, such as removing stop words, special characters, punctuations, and numbers, performing stemming, and then within each metric, tokenizing all reports belonging to the minority class and creating a dictionary of words. Then, we calculate the TF-IDF score of each word in this dictionary against all classes. In other words, if metric \textit{M} contains \textit{n}, different classes, every word in the dictionary is represented by \textit{n} TF-IDF score, each corresponds to an individual class in a vector of size \textit{n}. This vector is then normalized to re-scale the scores from 0 to 1. These normalized values represent the statistical relevance of a term belonging to a class, with values closer to 1 indicating higher importance. Since we aim only to identify minority classes using TF-IDF score vectors, we implement a filtering procedure based on a hyperparameter threshold to select words whose scores in minority class(es) are "significantly" higher than their scores in majority classes. As a result, every vulnerability description is defined by a set of \textit{n}-dimension TF-IDF vector if it shares any word with minority class word dictionary, where \textit{n} refers to the number of existing classes in metric \textit{M}. Then, we take the average of these vectors as the final TF-IDF vector representation of the input description. Fig. \ref{fig: TF-IDF_schema} shows the high-level overview of TF-IDF vector creation.

Let $\delta(M)$ be the dictionary of terms in minority class(es) in metric $M$:
\begin{equation}
    \delta(M) = \{w_1, w_2, ..., w_r\}
\end{equation}

If $M=\{c_1, ..., c_n\}$ represents the existing classes in metric $M$, for each $w \in \delta(M)$, we return $n$ scores (TF-IDF scores) corresponding to each $c \in M$:
\begin{equation}
    T(w, M)= [t^w_{c1}, ...,t^w_{cn}].
\end{equation}

Then we normalize each vector $T(w_i, M)$ by dividing all elements by the maximum value of the vector as $MAX[T(w_i, M)] + \epsilon$ ($\epsilon$ is a small number added to avoid divide by zero) and call it $T'(w_i, M)$.

Here, our target is to provide a secondary representation of the input text in addition to SecureBERT's output leveraging TF-IDF. This vector representation will be concatenated with SecureBERT's output and yield to the classification layer together. Therefore, given $T'(w, M)$, we first define a probability threshold $0<th<1$. Then,  for each $T'(w_i, M)$,
make sure its score corresponding to one potential value within a particular metric is greater than $th$ ($t^w_{i,fj} > th$) and the score corresponding to the other values is smaller than $1-th$, otherwise, remove the word from the dictionary.
In other words, we compare all \textit{n} the number of TF-IDF scores associated with each word with the threshold and identify those  whose value corresponding to a minority class is significantly higher than other classes (according to the threshold).  Any word that satisfies both conditions will be added to the signature term dictionary $Dict(M)$ corresponding to each metric. Therefore, $Dict(M)$ contains words that are highly informative about minority values within each metric.

For any training input CVE description within each metric prediction model, we check how many terms it shares with $Dict(M)$. Let $S=\{w'_1, ...,w'_n\}$ represent the terms in a CVE description, and $g(S, M)=S \cap Dict(M)$ be the set of shared terms between $S$ and $Dict(M)$. For any term $w'\in g(S, M)$, we retrieve the TF-IDF vectors using Eq. \ref{eq: tfidf_norm} and take the weighted average ($\overline{\mu}$)  of all vectors to construct the TF-IDF representation of the CVE:
\begin{equation}
    \mathcal{V}(S, M) = \overline{\mu}{\texttt{ }} (T'(w', M)), \forall w'\in g(S)
    \label{eq: tfidf_norm}
\end{equation}

$\mathcal{V}(S, M)$ is a vector with a size equal to the number of values within a metric representing the probability score of a CVE being associated with each value, with respect to the signature words. Since signature words are extracted from minor classes, the vector elements associated with the dominant class samples are quite often zero or close to zero, indicating that this vector represents the likelihood that a CVE is associated with a minor value, if the CVE shares any words with the signature word dictionary $Dict(M)$; otherwise, all elements would be zero.
Opting for the weighted average proves advantageous over selecting the maximum or minimum values when creating representative vectors. While our goal is to amplify the importance of the minority class when computing a CVE's TF-IDF representation, opting for the minimum value isn't suitable. Conversely, signature words lack uniqueness, implying their occasional appearance in other documents, even if infrequently, including those linked to dominant classes. As a result, assigning the maximum value could potentially associate with an erroneous class, yielding an inaccurate vector.

\begin{figure}[!ht]
 \center
  \includegraphics[width=8.5cm]{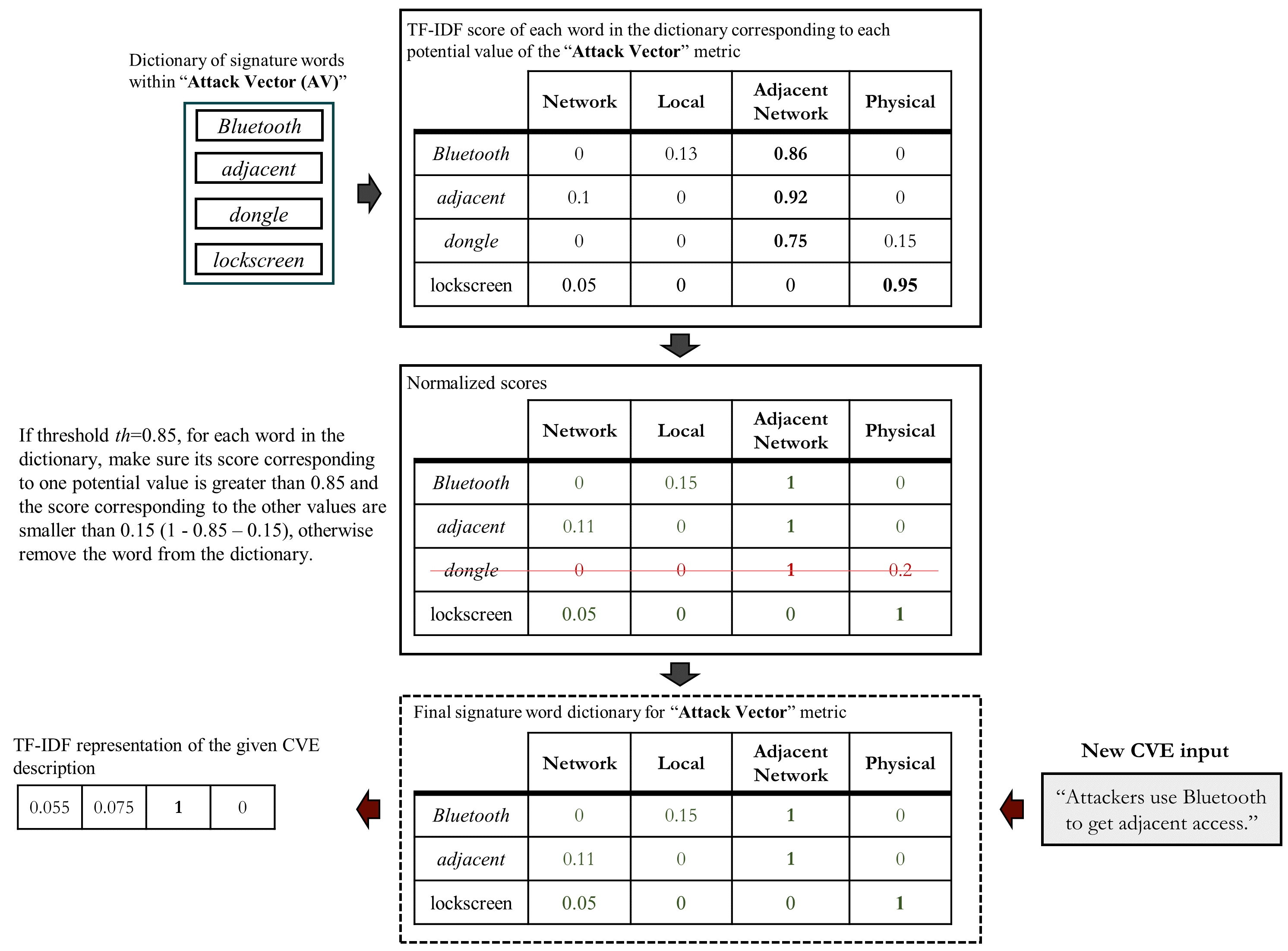}
  \caption{An example of TF-IDF module creation and usage.}
  \label{fig: TF-IDF_example}
\end{figure}

Fig. \ref{fig: TF-IDF_example} shows an example of how the TF-IDF works in the Attack Vector (AV) metric. Suppose \textit{Bluetooth}, \textit{adjacent}, \textit{dongle}, and \textit{lockscreen} are the top four most frequent words that exist in the minority values of AV, "Adjacent Network" and "Physical", that build the initial signature word dictionary. After collecting such frequent words, the TF-IDF score of each word corresponding to each potential value is calculated and then normalized. If threshold $th=0.85$, we exclude the word \textit{dongle} from this dictionary since it does not satisfy the two rules to keep the word in the dictionary as its score corresponding to the "Physical" value is $0.2$ which is larger than $1-th = 0.15$.
After generating the final dictionary, we look for the signature words in the new text input "\textit{Attackers use Bluetooth to get adjacent access.}", and extract the vectors associated with signature words  \textit{Bluetooth} and \textit{adjacent} and take the average to return the TF-IDF vector representation of the text, based on the created dictionary of the signature words.

\subsection{Evaluation}
In this section, we undertake a comprehensive assessment of our proposed model's performance in individual metric value prediction. Our evaluation is centered solely around CVE human-readable text. Furthermore, we furnish the experimental configurations necessary to replicate the model, and we delve into both the model's merits and shortcomings. Additionally, we present recommendations to enhance the efficacy of automating CVSS prediction.
\\

\noindent\textbf{Experimental Settings}\\
We used the NVD dataset that contains 28,090 CVE reports each assigned by a CVSS vector for training the proposed model. For each report, we create two vector representations, a vector generated by SecureBERT's tokenizer, and a customized TF-IDF vector utilizing the methodology discussed in the previous section. In addition, we conduct data resampling for minority class(es) within each metric by randomly duplicating samples, to reduce the impact of imbalanced data problem and avoid improper classification while training. We train the model through 10 epochs with a mini-batch size of 12 with a learning rate equal to $1e-5$. The training objective is to minimize the $Cross Entropy$ error using the $Adam$ optimizer. \\

\noindent\textbf{Model Performance}\\
Table \ref{tab: cvss_results} shows the comparative performance of our model in predicting every eight values in the CVSS base vector metrics, using a testing dataset of $5,357$ (20\% of the main dataset) CVE reports. We evaluated two versions of the model, one with and one without the TF-IDF module. This table shows the accuracy and the F-1 score at both micro and macro level, where micro calculates metrics in each case level by counting the total true positives, false negatives, and false positives, and macro evaluates metrics for each label, and find their unweighted mean. Considering the shortage of key information in CVE reports and the imbalance class problem, SecureBERT achieved $80\%-96\%$ prediction weighted accuracy, $80\%-95\%$ micro-level, and $73\%-91\%$ macro-level F1-score.

\begin{table*}
\centering
\scriptsize
 \begin{tabular}{|c |c |c| c | c | c |c | c |c | c |} 
  \hline

   Model & Metric & {AV} & {AC} & {PR} & {UI} & {S} & {C} & {I} & {A} \\
  \hline
SecureBERT & Accuracy & {\textbf{89.57}} & {\textbf{96.04}} & {\textbf{80.59}} & {\textbf{93.01}} & {\textbf{94.60}} & {83.23} & {84.35} & {\textbf{87.40}} \\

with & F1-Score (Micro) &{\textbf{89.45}} & {\textbf{95.96}} & {\textbf{80.58}} & {\textbf{93.01}} & {\textbf{94.39}} & {\textbf{83.23}} & {84.35} & {\textbf{87.04}} \\

TF-IDF & F1-Score (Macro) & {\textbf{76.16}} & {\textbf{81.10}} & {\textbf{73.67}} & {\textbf{91.35}} & {\textbf{89.91}} & {\textbf{80.70}} & {83.58} & {\textbf{72.47}} \\
\hline
 SecureBERT & Accuracy & {89.52} & {95.85} & {79.42} & {91.78} & {92.06} & {\textbf{83.25}} & {\textbf{84.99}} & {86.01} \\
 
 without & F1-Score (Micro)  &{89.38} & {95.84} & {79.42} & {91.78} & {92.06} & {83.21} & {\textbf{84.59}} & {85.91} \\
 
 TF-IDF  & F1-Score (Macro) & {75.54} & {80.34} & {72.03} & {89.21} & {86.88} & {80.37} & {\textbf{83.83}} & {71.41} \\
              

    \hline

\end{tabular}\hfil
     \caption{\label{tab: cvss_results} CVSS metric value prediction results}
\end{table*}

Based on the performance outcomes, the model incorporating the TF-IDF module demonstrates superior performance compared to the alternative model. This observation underscores the module's efficacy in accurately identifying the appropriate class, particularly when specific keywords are consistently present within corresponding classes.
Simultaneously, an initial examination of the distinction between micro and macro metrics underscores the influence of the unbalanced data issue, particularly in metrics characterized by a higher degree of imbalance. However, the prediction result of some metrics whose micro-macro performance gap is small, yet unbalanced, such as \textit{User Interaction (UI)} and \textit{Scope (S)}, shows the proposed model could successfully identify the correct class. This implies that the CVE reports typically provide the required information for these metrics. On the other hand, the high rate of false positives in both minor and dominating values of some metrics such as \textit{Confidentiality (C)} and \textit{Availability (A)} indicates the lack of key knowledge in CVE reports, leading to a lower rate of correct prediction.
\begin{figure}
\Center
\includegraphics[width=8cm]{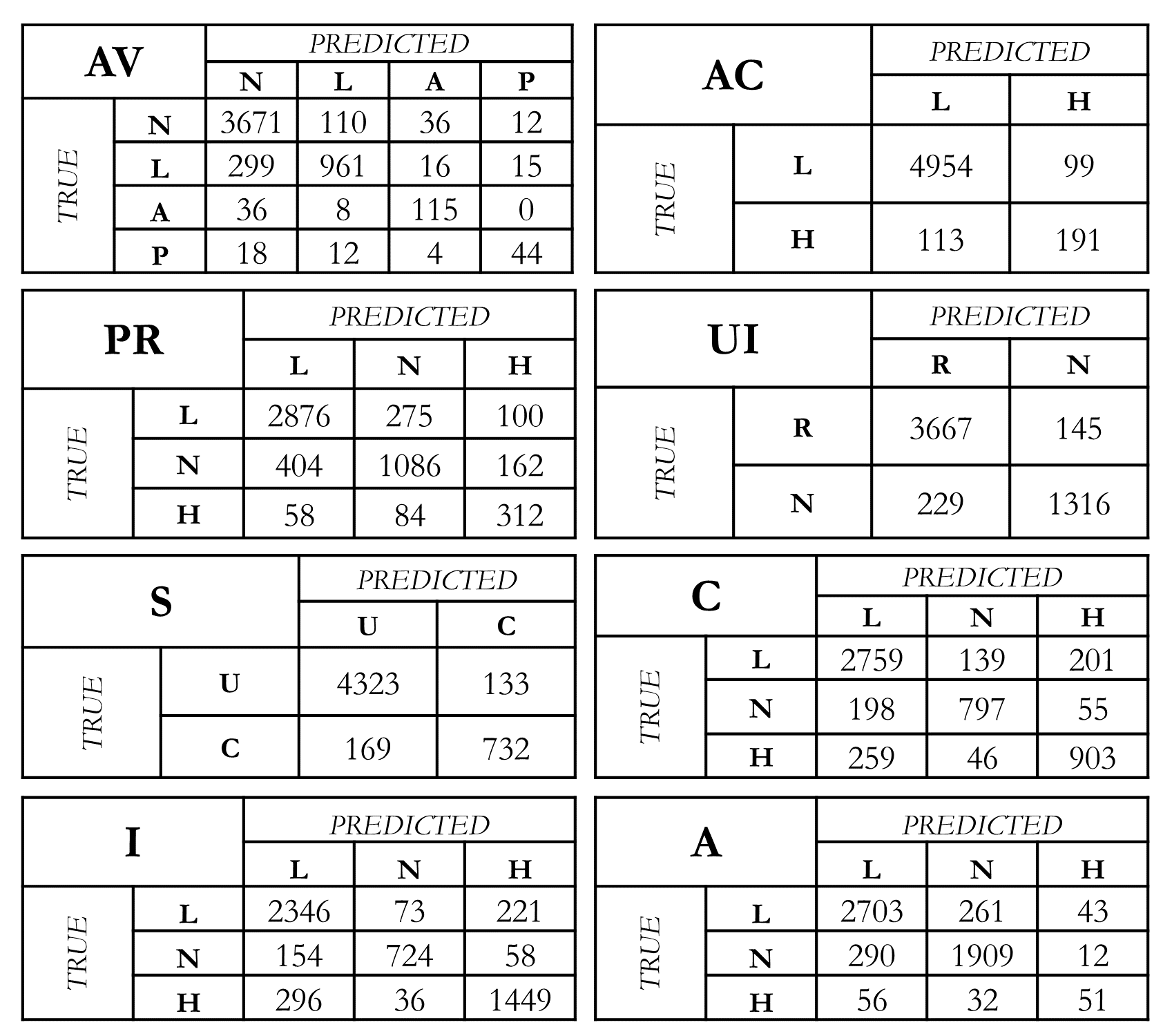}
\caption{Confusion matrix for the CVSS base metric prediction model.}
\label{tab:cvss_confmatrix}
\end{figure}
Table \ref{tab:cvss_confmatrix} shows the confusion matrix of each eight individual models.
\\
\\
\noindent\textbf{Shortages and Limitations}\\
As mentioned earlier, CVEs typically fail to provide all key information about the CVSS metrics. For example, Table \ref{tab: CVSS_misclass_example} shows the description of CVE-2012-1516 that is describing a vulnerability in VMware. The "Attack Complexity" metric describes "\textit{the conditions beyond the attacker's control that must exist in order to exploit the vulnerability. Such conditions may require the collection of more information about the target, the presence of certain system configuration settings, or computational exceptions}" \footnote{\url{https://www.first.org/cvss/v3.0/specification-document}}. This is a pretty deep and conceptual definition requiring some background knowledge and fine-grain information about the vulnerability to specify if the value of this  metric (Attack Complexity) for the given CVE is low or high.
According to the vendor report, the basic requirement for this attack to exploit is 4GB of memory, and those with less than 4GB of memory are not affected, hence the "Attack Complexity" for this CVE is rated as "Low".
The CVE description (see Table \ref{tab: CVSS_misclass_example}) failed to provide such detailed information (i.e., the 4GB RAM need), making it pretty obscure even for an expert to predict the metric value based on the provided text. 

\begin{table*}
\centering
 \begin{tabular}{|l | l |} 
\hline 
 \multicolumn{2}{p{\textwidth-2cm}}{\textbf{CVE-2012-1516}: \textit{The VMX process in VMware ESXi 3.5 through 4.1 and ESX 3.5 through 4.1 does not properly handle RPC commands, which allows guest OS users to cause a denial of service (memory overwrite and process crash) or possibly execute arbitrary code on the host OS via vectors involving data pointers.}} \\
  \hline

   \multicolumn{1}{|l|}{\textbf{Attack Complexity}} & \multicolumn{1}{|p{10cm}|}{High} \\
   \hline
  \multicolumn{1}{|l|}{\textbf{Reason}} &
  \multicolumn{1}{|p{13cm}|}{The only required condition for this attack is for virtual machines to have 4GB of memory. Virtual machines that have less than 4GB of memory are not affected} \\
   \hline
\end{tabular}\hfil
     \caption{\label{tab: CVSS_misclass_example} Shows an example of an uninformative CVE description for the purpose of "Attack Complexity" metric prediction.}
\end{table*}

CVE database comprises many similar reports making both manual or automated CVSS perdition hard or even infeasible, solely based on the description.  Some works leveraged other standards such as CWEs to enrich the CVE texts for better prediction. However, this is highly problematic and leads to inconsistency since such standards represent the threats with a high-level view and may not provide fine-grain information about the vulnerability's detailed properties.  
An additional strategy to enhance automated CVSS prediction involves leveraging external resources like vendor vulnerability reports. These reports often furnish in-depth insights into vulnerabilities, encompassing various dimensions such as impacts, platforms, permissions, techniques, tactics, and procedures. To capitalize on this, a robust text analytics tool such as SecureBERT can be employed. By processing the contextual information and scrutinizing semantic connections, this tool holds the potential to extract specific actions and attributes, thereby representing each metric more effectively assuming the input text sufficiently provides the requisite information.
\\
\\
\noindent\textbf{Comparison with ChatGPT}\\
As noted before, although the recent large language models have demonstrated reasonable performance in accomplishing general text analytics tasks, it tends not to perform well in domain-specific and fine-grain cybersecurity tasks. Unlike ChatGPT, our model maintains continuous learning and refinement to always obtain high accuracy of the vulnerability assessment via the automated CVSS predictor. This is essential as predicting the CVSS score accurately requires continuous updates and regular model adjustments as new attack vector information emerges. 
In order to evaluate ChatGPT's predictive capabilities for CVSS vectors, we conducted a comparative analysis. For this purpose, we utilized a set of 100 random CVEs that were disclosed in 2023. These CVEs had not been encountered by either our model or ChatGPT prior to the latter's knowledge cutoff in 2021. Subsequently, we tasked ChatGPT with predicting the corresponding CVSS vectors for these CVEs, enabling a comparison with our own model's predictions. The results, as presented in Table \ref{tab: cgpt_cvedrill}, demonstrate a notable disparity in performance between the two models. Our model exhibited a considerably higher accuracy in predicting 7 out of 8 vectors, whereas ChatGPT struggled, highlighting its shortcomings in this context.
\begin{table}[]
\scriptsize
\begin{tabular}{l|llllllll|}
\cline{2-9}
{\color[HTML]{000000} }                                           & \multicolumn{8}{c|}{{\color[HTML]{000000} Prediction Accuracy}}                                                                                                                                                                                                                                                                                                                                                                                                          \\ \hline
\multicolumn{1}{|l|}{{\color[HTML]{000000} \textbf{CVSS vector}}} & \multicolumn{1}{l|}{{\color[HTML]{000000} \textbf{AV}}}   & \multicolumn{1}{l|}{{\color[HTML]{000000} \textbf{AC}}}   & \multicolumn{1}{l|}{{\color[HTML]{000000} \textbf{PR}}}   & \multicolumn{1}{l|}{{\color[HTML]{000000} \textbf{UI}}}   & \multicolumn{1}{l|}{{\color[HTML]{000000} \textbf{S}}}    & \multicolumn{1}{l|}{{\color[HTML]{000000} \textbf{C}}}    & \multicolumn{1}{l|}{{\color[HTML]{000000} \textbf{I}}}    & {\color[HTML]{000000} \textbf{A}}    \\ \hline
\multicolumn{1}{|l|}{{\color[HTML]{000000} ChatGPT}}              & \multicolumn{1}{l|}{{\color[HTML]{000000} 0.52}}          & \multicolumn{1}{l|}{{\color[HTML]{000000} 0.78}}          & \multicolumn{1}{l|}{{\color[HTML]{000000} \textbf{0.85}}} & \multicolumn{1}{l|}{{\color[HTML]{000000} 0.55}}          & \multicolumn{1}{l|}{{\color[HTML]{000000} 0.43}}          & \multicolumn{1}{l|}{{\color[HTML]{000000} 0.36}}          & \multicolumn{1}{l|}{{\color[HTML]{000000} 0.81}}          & {\color[HTML]{000000} 0.42}          \\ \hline
\multicolumn{1}{|l|}{{\color[HTML]{000000} CVEDrill}}             & \multicolumn{1}{l|}{{\color[HTML]{000000} \textbf{0.82}}} & \multicolumn{1}{l|}{{\color[HTML]{000000} \textbf{0.86}}} & \multicolumn{1}{l|}{{\color[HTML]{000000} 0.76}}          & \multicolumn{1}{l|}{{\color[HTML]{000000} \textbf{0.91}}} & \multicolumn{1}{l|}{{\color[HTML]{000000} \textbf{0.87}}} & \multicolumn{1}{l|}{{\color[HTML]{000000} \textbf{0.81}}} & \multicolumn{1}{l|}{{\color[HTML]{000000} \textbf{0.81}}} & {\color[HTML]{000000} \textbf{0.77}} \\ \hline
\end{tabular}
\caption{Comparative evaluation of predictive accuracy for CVSS vectors in a sample of 100 randomly selected CVEs disclosed in 2023.}
\label{tab: cgpt_cvedrill}
\end{table}

Table \ref{Tab: chatgpt_cvss} in Appendix shows an overview of ChatGPT's predictive performance in estimating CVSS vectors for ten vulnerabilities we chose randomly from the recently disclosed CVEs. 


\section{Automated CVE to CWE Classification} \label{sec: Automated CVE to CWE Classification}
Cyber attacks enable malicious actors to break intended security policies by circumventing protective systems or influencing system resources or behavior. As a result, the attack leads to behavior that violates the victim's intended security regulations. Typically, attackers exploit a vulnerability by abusing an existing weakness in a system.
CWE is a hierarchically designed dictionary of software weaknesses for the purpose of understanding software flaws, their potential impacts if exploited, and identifying means to detect, fix, and prevent errors.
CWE classes are organized hierarchically, with higher-level classes providing higher-level attack characteristics and lower-level classes inheriting the parent classes' features and adding micro-granularity details corresponding to the potential threats. Therefore, determining the best path from a root to lower-level nodes allows for the acquisition of fundamental and functional directions for detecting and analyzing the different properties of a vulnerability.

For example, given 'CVE-2004-0366: A \textit{SQL injection vulnerability} in the libpam-pgsql library before to 0.5.2 allows attackers to execute arbitrary SQL statements.', the description captures the attack action (execute arbitrary SQL statements) within a specific object (libpam-pgsql library) and specifies the consequence (SQL injection). While this low-level, product-oriented description depicts SQL injection exploitation, it falls short of clearly defining the characteristics of this malicious behavior, which is highly required to address possible prevention and/or detection measures. The supplementary CWE (CWE-89: SQL Injection) \footnote{\url{https://cwe.mitre.org/data/definitions/89.html}} gives high-level and non-product-specific information by addressing three critical questions: (1) why the attack is used: the system does not properly neutralize special elements; (2) how the attack is used: by changing the intended SQL query; and (3) what the probable results are: access or modify application data; and bypass protection mechanism.

The above-mentioned case is a confirmatory example to show how a CWE can paint a clear picture of the existing holes in the systems and reveals potential factors leading to vulnerability exploitation. Obtaining these factors is closely associated with the paradigm of pinpointing applicable mitigation or detection methods. For example, we can apply an "accept known good" input validation strategy, i.e., using a set of legit inputs that strictly 
conform to specifications and rejects the rest, to mitigate SQL injection. Besides, we can detect SQL injection by performing an automated static analysis (e.g., bytecode or binary weakness analysis), dynamic analysis (e.g., database or web service scanners), or design review (e.g., formal methods). 

\begin{figure}[ht]
\Center
\includegraphics[width=6cm]{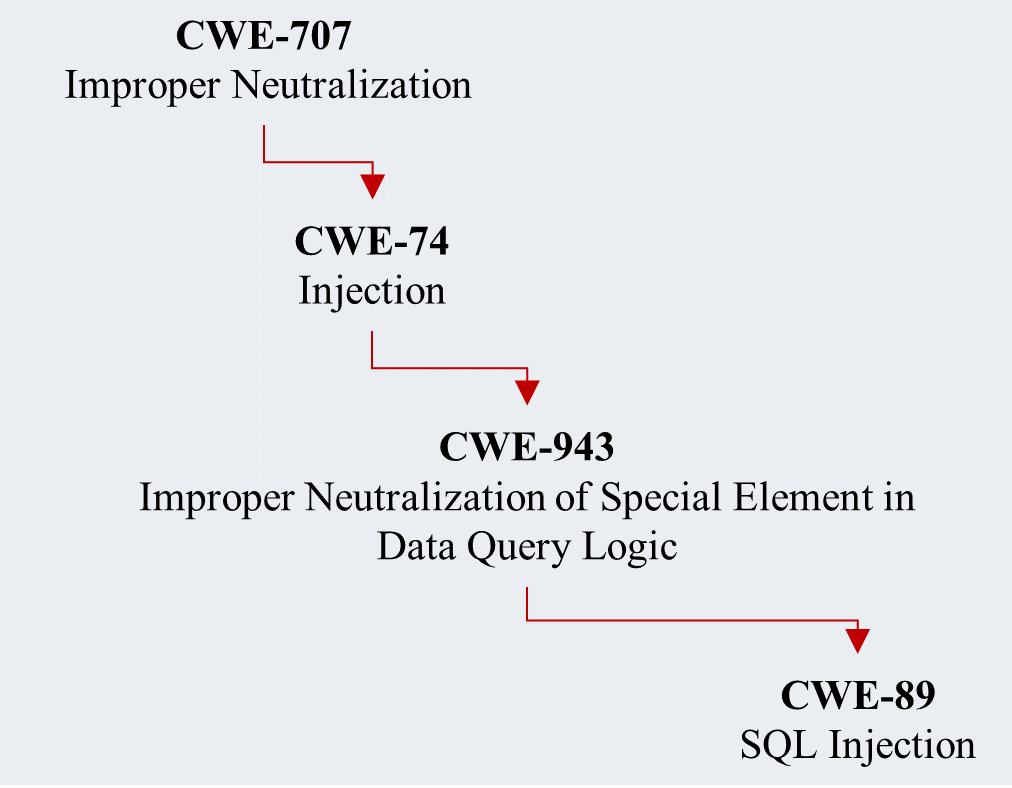}
\caption {This figure depicts the hierarchical representation of the CWEs. The red boxes show the CWE-89's relatives in the higher levels. This hierarchy plays an important role in understanding the character of the weaknesses in different levels of detail.} 
\label{fig:cwesnap}
\end{figure} 

Fig. \ref{fig:cwesnap} shows the hierarchical representation of the CWEs. Analyzing the path from the root all the way to any node in the lower levels is indispensable since each node reveals different functional directions to learn a weakness. For example, by tracking the path from the root node, CWE-707, to  CWE-89, we realize that the SQL injection (CWE-89) is a result of an improper neutralization of special elements in data query logic (CWE-943), where both weaknesses are associated with injection (CWE-74), and the injection itself is the result of the improper formation and neutralization of a message within a product before it is read from an upstream component or sent to a downstream component (CWE-707). Incorporating this thorough knowledge graph helps to maintain countermeasures from different senses, even if the most fine-grain node is not available. For example, assume that only two coarse-grain candidates in different levels of hierarchy, CWE-707, and CWE-74, are available for CVE-2004-0366, while the most fine-grain weakness (CWE-89) is not discovered yet. Although fine-grain SQL injection characteristics are not exposed, investigating the coarse-grain candidates helps to find the common consequences and impacts, and accordingly extract defense actions against improper neutralization and injection (e.g., filtering control-plane syntax from all input). This example explicitly highlights the significance of the existing hierarchical structure of CWEs and shows how useful it is in perceiving defense actions. 
A significant number of CVEs are currently mapped to a small set of CWE classes. Currently, about 70\% of the CWE classes have fewer than 100 CVEs and about 10\% have no CVEs mapped to them, and only 10\% have more than 500 CVEs. 

\subsection{Challenges}
Associating CVEs with CWEs allows cybersecurity researchers to understand the means, assess the impact, and develop solutions to mitigate threats. However, the problem is loaded with challenges. A CVE can be mapped to multiple and interdependent CWEs on the same route, leading to uncertainty. On the other hand, high-quality mapping information is scarce since CVEs are currently mapped manually to CWEs, which is neither scalable nor reliable. Manual mapping of CVEs is not a practical strategy since new CVEs are introduced at a quick rate. Hence, effective approaches for automating the mapping of CVEs to CWEs are critical for addressing ever-increasing cybersecurity risks.

CVEs are typically short and low-level descriptions written in an advanced language. The format and terminology used in these reports require cybersecurity knowledge and a deep context understanding for automation. Traditional text mining approaches that work based on word frequency fail to adequately capture the context and the semantic relationships between the words. Additionally, the modern NLP tools and pre-trained language models which are trained only on general English corpus with no specific focus on cybersecurity, may not effectively and optimally process such advanced language. Therefore, a domain-specific model that is trained on cybersecurity data can help in more efficient.

\begin{figure*}
\Center
\includegraphics[width=0.8\textwidth]{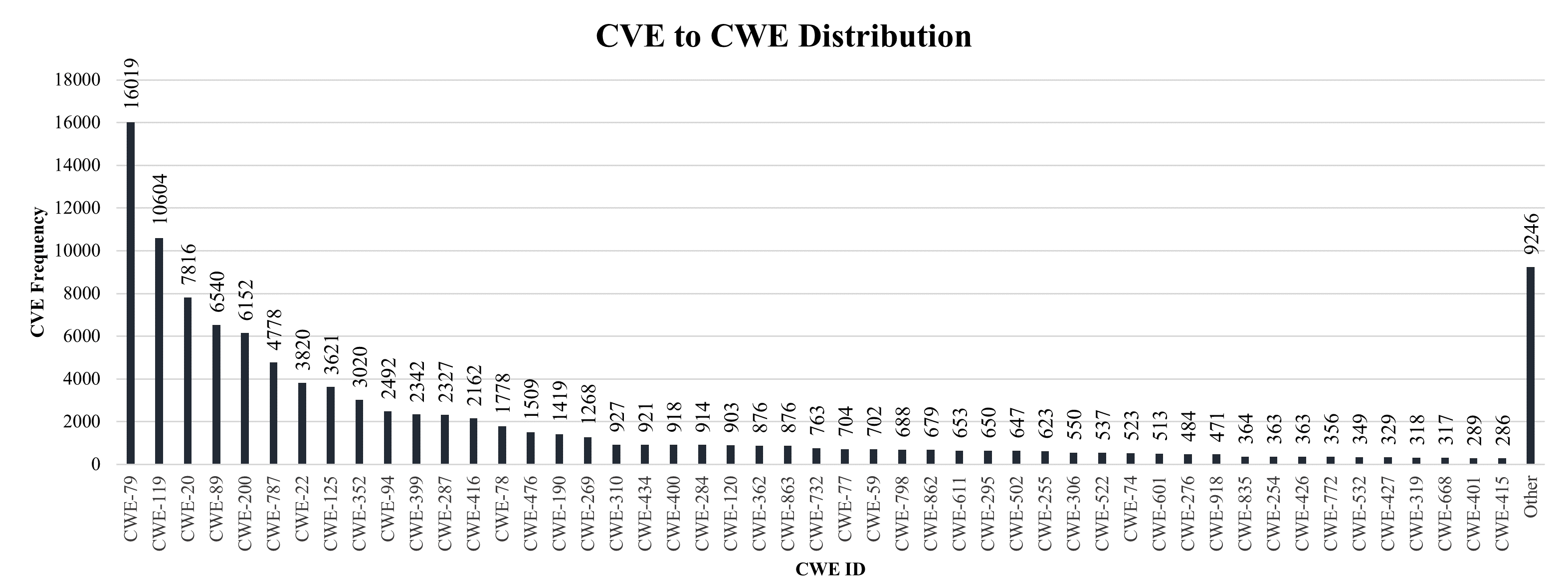}
\caption {Shows the distribution of common CWEs } 
\label{fig:cwedist}
\end{figure*}

Furthermore, the available CVE-to-CWE dataset is highly unbalanced meaning that some CWEs are much more common than others. A portion of this issue is justified, as NVD focuses on more general CWEs that encompass a broader range of specific characteristics, rather than using several distinct CWEs. For example, "CWE-79: Cross-site Scripting (XSS)" is the most frequently reported vulnerability in the NVD classification, accounting for 16,019 CVE reports. Meanwhile, NVD has only used a variant of the XSS vulnerability, the "CWE-87: Improper Neutralization of Alternate XSS Syntax" only one time. NVD's similar use of weaknesses to represent CVEs demonstrates that several primary CWEs cover specific properties and can thus represent a large group of similar CWEs. 
However, this does not apply to all CWEs as some of them are not commonly exploited, yet carry unique properties. Therefore, there must be a robust and well-defined approach to be able to classify CVEs to uncommon CWEs appropriately. Table \ref{fig:cwedist} shows the distribution of the top 50 most commonly used CWEs by NVD. As depicted, the appearance of a few CWEs such as "CWE-79: Cross-site Scripting (XSS)", "CWE-119: Improper Restriction of Operations within the Bounds of a Memory Buffer", and  "CWE-20: Improper Input Validation" is much higher than other weaknesses.

\subsection{Methodology}
We addressed the issue of automatic CVE to CWE classification as a means of fine-tuning SecureBERT. We created a hierarchical classification layer on top of SecureBERT as a downstream task. This model takes a CVE description as input and returns corresponding CWE classes at each level of the hierarchy. To this end, we developed a top-down strategy, training a classifier on each node (class) in the CWE hierarchical tree. The classifier makes a choice for each node on whether to belong to a different sibling. The primary goal of this hierarchical structure is to direct the model's attention to the commonalities and contrasts between the sibling nodes.
This helps in lowering the model complexity by minimizing the number of output classes during the training procedure, as well as boosting model accuracy by allowing each classifier to focus on its children without being confused with other siblings' children. 

\begin{figure}[!ht]
\centering
\includegraphics[width=4cm]{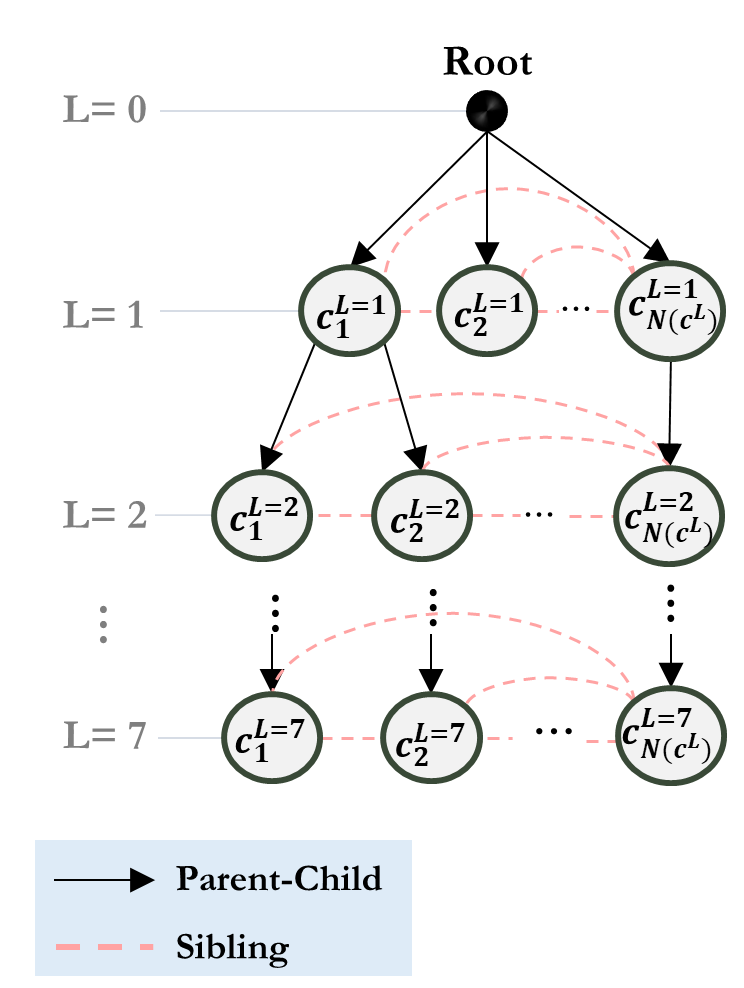}
\caption {Shows the CWE tree structure.} 
\label{fig:CWE_tree}
\end{figure}
\begin{figure*}[ht]

  \subfloat[CVE to CWE hierarchical model.]{
  \label{fig: cve2cwe_model}
	\begin{minipage}[c][0.95\width]{
	   0.5\textwidth}
	   \centering
	   \includegraphics[width=6.2cm]{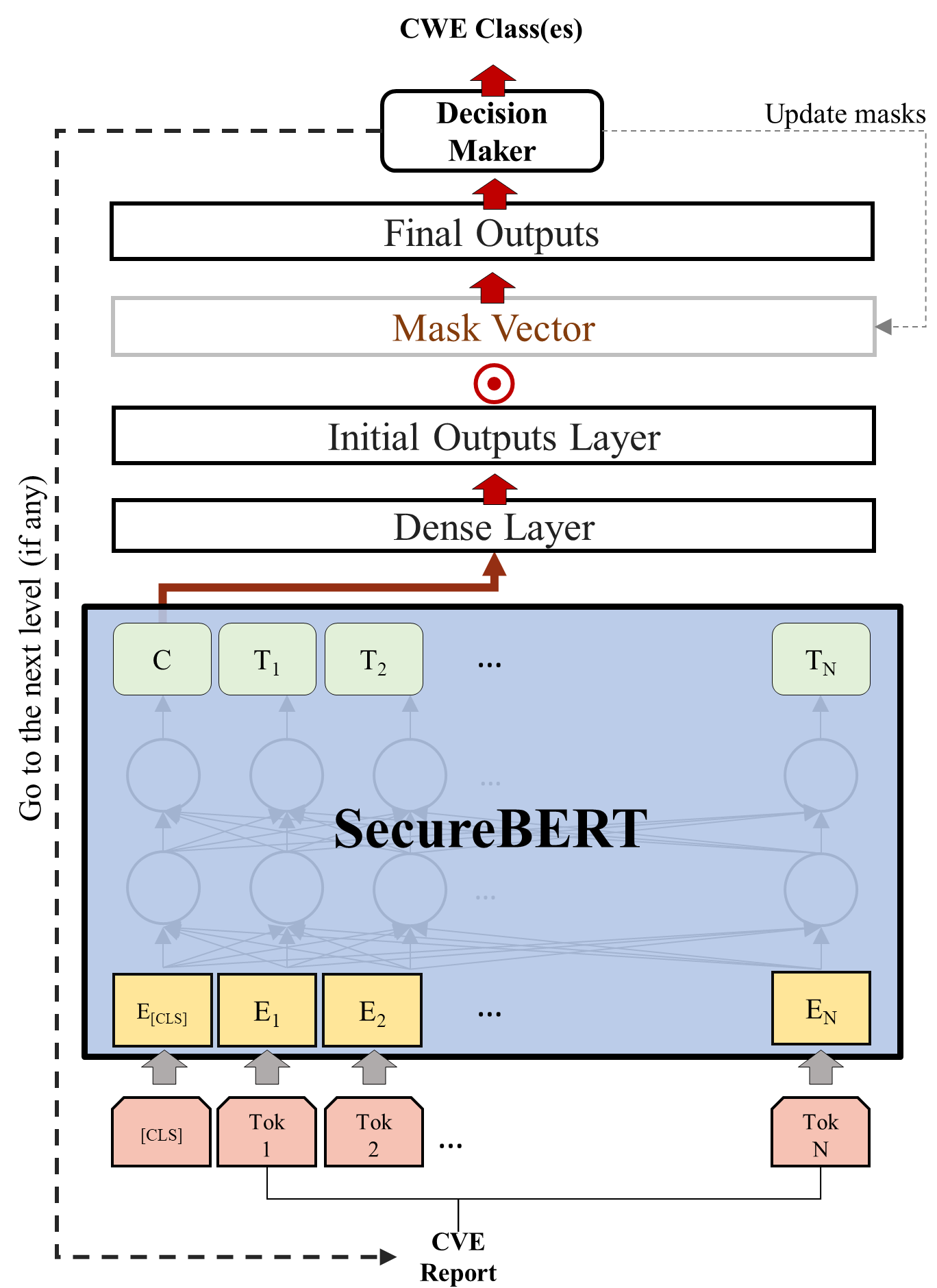}
	\end{minipage}}
 \hfill 	
  \subfloat[CVE to vulnerability type model.]{
  \label{fig: cve2vt_model}
	\begin{minipage}[c][0.8\width]{
	   0.5\textwidth}
	   \centering
	   \includegraphics[width=6cm]{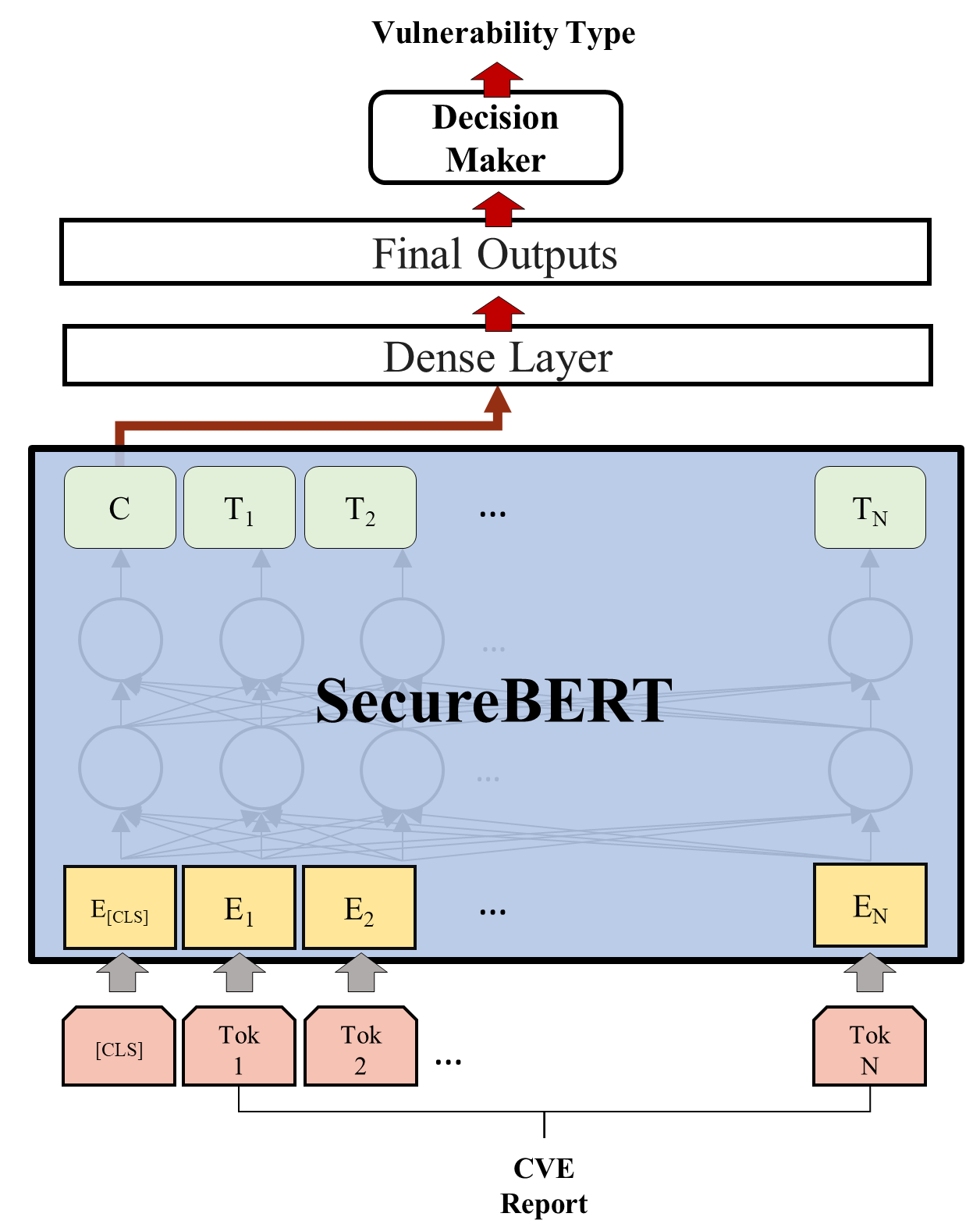}
	\end{minipage}}
 \hfill	
\caption{Show the model architectures for classifying CVEs to }
\end{figure*}

 
Fig. \ref{fig:CWE_tree} shows the CWE's tree structure in advance and demonstrates the existing levels, and "parent-children" and "sibling" relationships between the nodes. Tracing this tree from left, let's denote the $i_{th}$ CWE at Level $L$ by $c_i^{L}$ where $i\in\{1,...,N(c^{L})\}$ and $L\in \{1, ..., 7\}$. In our model design, for any given CVE, we aim to start predicting the associated CWEs from the $L=1$ and trace the CWE hierarchical tree for the correct class at each level to the node in the lowest possible level. Let's denote each CWE as a node $c_i^{L}$, and the total number of CWEs at level $L$ with $N(c^{L})$. Let $G(c_i^{L})$ represents all the children of $c_i^{L}$:

The model initially classifies the given CVE to one or more $c_i^{L}$ and then as a next step, it only focuses on the children of the predicted parent $\textbf{G}(c_i^{L})$ and moves forward until it reaches the most fine-grained level possible (or desired).
We use the pre-trained SecureBERT as the base model and add a classification layer on top which trains hierarchically. As depicted in Fig. \ref{fig: cve2cwe_model}, the model takes the CVE description tokenized by SecureBERT's tokenizer and passes it through the transformers stack, and returns the embedding vector of the first token ([\textit{CLS}] token). This hidden state vector [\textit{CLS}] is an aggregate representation of the entire input used for classification tasks.
This vector is connected to a pooling layer followed by a dense layer of size $N(c^L)$ returning the logits with no activation function, where $N(c^L)$ represents the total number of nodes in layer $L$. 

In order to make the model work in a hierarchical form, we multiply the output logits with a given masking vector $MV$, and forward the results (final logits) to the decision-making component. 
$MV$ is a sparse vector whose size equals to the initial output layer ($N(L+1)$) generated based on the classified node in the previous level representing the classified node's children. 
Assume, for an input CVE, the model classifies it to node $c_i^{L}$. If the total number of nodes (aka, the total number of possible CWE classes) in the next level $L+1$ equals $N(L+1)$, $MV$ would be a vector of $0$s and $1$s in which, $1$s represent the indices of $c_i^{L}$'s children nodes in the level $L+1$, and $0$s refer to the non-children nodes. The initial value of $MV$'s elements in level $L=1$ would be all $1$s since in the first round of training, there are only parents and children are defined. 
We multiply this vector to the initial output layer using element-wise multiplication to help the model recognize which nodes it should be focusing on during training and backpropagation. In other words, this multiplication turns off the nodes that are not the children of previously classfied classes, reducing complexity by lowering the number of potential classes for each input.

In addition to the hierarchical classification, we also consider more general groups of CWEs, so-called vulnerability types. Following the MITRE guideline \footnote{\url{https://github.com/center-for-threat-informed-defense/attack_to_cve/blob/master/methodology.md}}, similar weaknesses tend to share identical attack procedures and consequently exhibit similar characteristics, wherein this resemblance pertains to the CVEs linked with akin CWE classifications. The MITRE guideline outlines 27 unique vulnerability categories.
As previously stated, CWEs have two sorts of relationships including hierarchical (parent-children) and lateral (sibling). We leverage such relationships and manually map CWEs to their corresponding vulnerability types. Then, we build an automated model to classify CVEs to such vulnerability types. This classification holds significant importance since these vulnerability classifications are linked to MITRE ATT\&CK techniques, forming a crucial linkage for threat intelligence, evaluating risks, analyzing security gaps, and implementing countermeasures.

\begin{table*}[ht]
    \Centering

    \scriptsize
    \begin{tabular}{|l |p{4cm}|p{3cm}|p{6cm}|}
    \hline
        ID & \Centering{\textbf{Vulnerability Type}} & \Centering{\textbf{CWE ID(s)}} & MITRE ATT\&CK Techniques \\ \hline
        
       1 & General Improper Access Control & 284, 285, 287, 862, 863 &   \\ \hline
        
       2 & Improper Restriction of Excessive Authentication Attempts & 306, 307 & T1078 (Valid Accounts) / T1110.001 (Brute
Force: Password Guessing)\\ \hline
        
       3 & Authentication Bypass by Capture-replay & 294 & T1190 (Exploit Public-Facing Application)
/ T1040 (Network Sniffing) \\ \hline
        
       4 & Overly Restrictive Account Lockout Mechanism & 645 & T1446 (Device Lockout) / T1531 (Account Access
Removal) / T1110 (Brute Force\\ \hline
        
       5 & Use of Password Hash Instead of Password for Authentication & 836 & T1550.002 (Use Alternate Authentication Material:
Pass the Hash \\ \hline
        
       6 & General Credential Management Errors & 255, 256, 257, 260, 261 & T1552 (Unsecured Credentials) / T1078 (Valid
Accounts) \\ \hline
        
       7 & Cleartext Transmission of Sensitive Information & 319 & T1552 (Unsecured Credentials) / T1078 (Valid
Accounts) / T1040 (Network Sniffing)\\ \hline
        
       8 & Hard-coded Credentials & 798 & T1078.001 (Default Accounts)\\ \hline
        
      9 &  Weak Password/Hashing & 328, 916 & T1078 (Valid Accounts) / T1110 (Brute Force)\\ \hline
        
      10  & General Cryptographic Issues & 310, 324, 325, 326 & T1078 (Valid Accounts) / T1110 (Brute
Force) / T1557 (Man-in-the-Middle) / T1040
(Network Sniffing)\\ \hline
    11    & XML External Entity (XXE) & 611, 776 & T1059 (Command and Scripting Interpreter)
/ T1005 (Data from Local System) / T1046 (Network Service Scanning) \\ \hline
        
      12  & XML Entity Expansion (XEE) & 776 & T1499.004 (Endpoint Denial of Service: Application or System Exploitation \\ \hline
        
    13    & URL Redirection to Untrusted Site ('Open Redirect') & 601 & T1036 (Masquerading) / T1566.002 (Phishing:
Spearphishing Link)
\\ \hline
        
      14  & Cross-site Scripting (XSS) & 79, 692 & T1059.007 (Command and Scripting Interpreter:
JavaScript/JScript) / T1557 (Man-in-the-middle Browser) / T1189 (Drive-by Compromise) / T1204.001 (User Execution: Mali-
cious Link)\\ \hline
        
    15    & OS Command Injection & 78 & T1059 (Command and Scripting Interpreter)
/ T1133 (External Remote Service) \\ \hline
        
      16  & SQL Injection & 89, 564 & T1059 (Command and Scripting Interpreter)
/ T1005 (Data from Local System), T1505.003
(Server Software Component: Web Shell), T1136
(Create Account) / T1190 (Exploit Public-Facing
Application) / T1565.001 (Data Manipulation) \\ \hline
        
    17   &  Code Injection & 94 & T1059 (Command and Scripting Interpreter) \\ \hline
        
      18  & Directory Traversal (Relative and Absolute) & 20, 22, 23, 24, 25, 26, 27, 28, 29, 30, 31, 32, 33, 34, 35, 36, 37, 38, 39, 40 & T1202 (Indirect Command Execution)\\ \hline
        
    19    & Symlink Attacks & 59, 61, 62, 64, 65, 73, 363 & T1202 (Indirect Command Execution)\\ \hline
        
      20  & Untrusted/Uncontrolled/Unquoted Search Path & 426, 427, 428 & T1574 (Hijack Execution Flow) \\ \hline
        
    21    & Unrestricted File Upload & 434, 351, 436, 430, 73, 183, 184 & T1505.003 (Server Software Component: Web
Shell) / T1059 (Command and Scripting Interpreter) \\ \hline
        
      22  & Deserialization of Untrusted Data & 502 & T1059 (Command and Scripting Interpreter)\\ \hline
        
    23    & Infinite Loop & 835 & T1499.004 (Endpoint Denial of Service: Application or System Exploitation) \\ \hline
        
      24  & Cross-site Request Forgery (CSRF) & 352 & T1068 (Exploitation for Privilege Escalation)
/ T1204.001 (User Execution: Malicious Link \\ \hline
        
    25    & Session Fixation & 384 & T1563 (Remote Service Session Hijacking\\ \hline
        
      26  & Uncontrolled Resource Consumption & 400, 664, 770, 771, 779, 920, 1235, 410 & T1499 (Endpoint Denial of Service)\\ \hline
        
    27    & Server-Side Request Forgery (SSRF) & 918 & T1090 (Proxy) / T1135 (Network Discovery)
/ T1005 (Data from Local System) / T1133 (External Remote Service \\ \hline
        
    \end{tabular}
    \caption{Mapping CWEs to vulnerability types.}\label{tab: cwe2vtype}
\end{table*}

Table \ref{tab: cwe2vtype} This figure demonstrates our manual effort in mapping CWE to vulnerability types, as well as MITRE ATT\&CK techniques associated with each type. Vulnerability types are mainly defined for the purpose of finding CVEs with a common set of techniques used to exploit that is useful in mapping CVEs to MITRE ATT\&CK techniques. Therefore, such vulnerability types are not comprehensive and do not cover all existing CWEs.

Fig. \ref{fig: cve2vt_model} shows the CVE-to-VT classification model architecture. It follows a similar structure as the hierarchical model with two main differences. First, it is a flat model, and therefore, there is no masking vector. In addition, in the decision-making component, we utilize Softmax to maximize the probability of the predicted output. In short, the output size of this multiclass classification model is the fixed number $27$ which equals the total number of vulnerability types defined in Table \ref{tab: cwe2vtype}.

\subsection{Evaluation}
In this section, we undertake a comparative evaluation of our proposed model's performance in classifying CVEs to CWEs using only human-readable language from CVEs.
We train two different classification models, a hierarchical CVE to CWE classification and a CVE to vulnerability types classification model, and perform a comparative evaluation on each.
Additionally, we provide experimental settings for reproducing the model and discuss the model's strengths and drawbacks.

\subsection*{Hierarchical CVE to CWE Classification Model}
For training the hierarchical model, we use a list of 63,481 CVE descriptions and classified them to one or more CWEs at each level of the hierarchy. Note that, the lowest level CWE used by NVD is $L=5$, and therefore, we separately train five models corresponding to each level. 
For each model, we use Binary Cross Entropy (BCE) with logits as a loss function for this multiclass multi-label classification task and evaluate by the hit rate and F1-score.  We use $10$ epochs with Adam optimizer and the initial learning rate of $3e-5$. For training each model at level $L+1$, we further fine-tune the trained model in level $L$ instead of the raw pre-trained SecureBERT, to ensure it conveys the most recent information about the CVE texts. In the decision-making component in the output layer of the model, we perform min-max normalization to produce a probability score corresponding to each output unit.

The NVD dataset had classified 121,768 CVEs to 296 CWEs as of the date of authoring this article. However, there is a significant disparity in the frequency with which CWEs are used, indicating that some CWEs are significantly more prevalent than others. This implies that infrequent CWEs are either rare or have characteristics in common with one of the frequent CWEs. 

\begin{figure}
 \center
  \includegraphics[width=4cm]{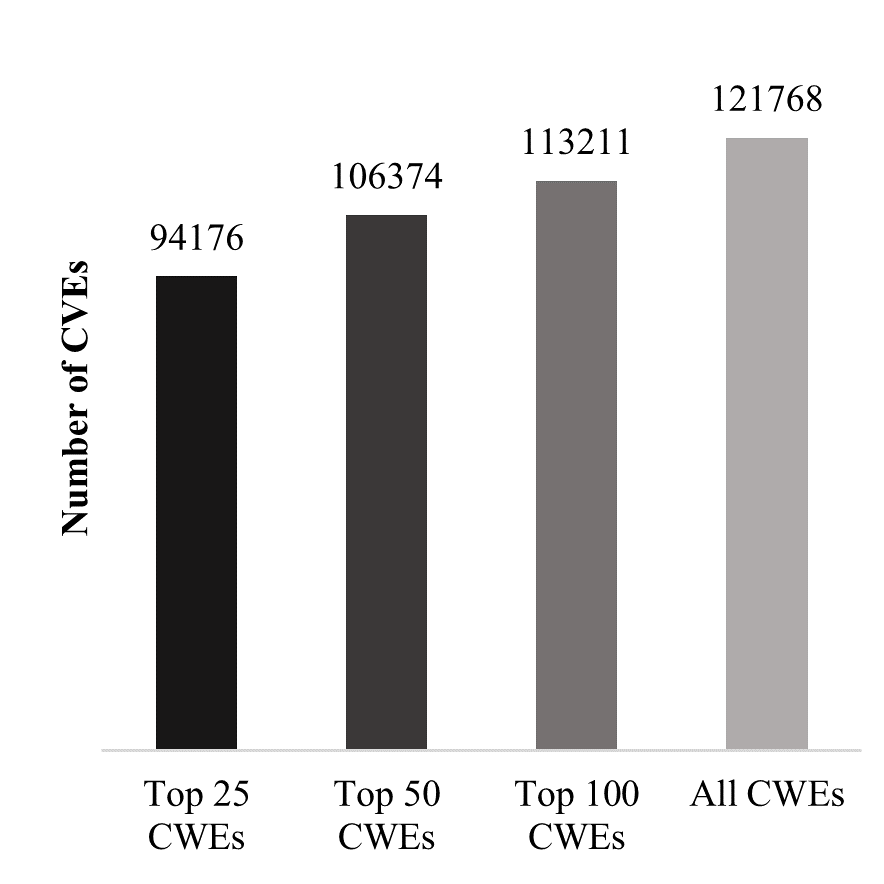}
  \caption{Distribution of CWEs in CVE to CWE classification.}
  \label{fig: TopCWE_dist}
\end{figure}

As depicted in Table \ref{fig: TopCWE_dist}, $77.3\%$ ((94,176 CVEs)), $87.4\%$ (106,374 CVEs), and $93\%$ (113,211 CVEs) of the CVEs have been classified to top 25, 50, and 100 CWEs, respectively. Note that, in calculating true predictions, if the instance was originally labeled by more than one CWE (multilabel instances), and the model catches one with no false positives, it is considered correct classification because any correct prediction can lead to the correct label in the lower levels.

In the hierarchical model, we classify CVEs to the common CWEs and evaluated the model in different stages. Table \ref{tab: cve2cwehir_eval} shows the performance of the hierarchical model in classifying CVEs as the top 25,  50, and 100 most frequent CWEs. According to this table, the success rate of our proposed model ranges from $90.77\%$ to $97.51\%$ in terms of hit rate (accuracy), and $84.91\%$ to $96.08\%$ in terms of F1-score, concerning the CWE space and the level of hierarchy. Such results demonstrate a higher success rate in classifying CVEs belonging to the top 25 CWEs due to the higher number of samples available for training the model.
\begin{table*}[!ht]
\centering
 \caption{SecureBERT performance evaluation on the hierarchical classification of CVEs to top 25, 50, and 100 CWEs where HR indicates the hit rate (accuracy in this case) and F1 refers to the F1-score.}
 \label{tab: cve2cwehir_eval}

\begin{tabular}{|l|c|c|c|c|c|c|c|c|c|}
 \hline
 \multicolumn{1}{|c|}{}& \multicolumn{3}{c|}{\textbf{Top 25 CWEs}} & \multicolumn{3}{c|}{\textbf{Top 50 CWEs}} & \multicolumn{3}{c|}{\textbf{Top 100 CWEs}} \\
 \hline
 L & No. CWEs &\textbf{HR} & \textbf{F1} & 
 No. CWEs &\textbf{HR} & \textbf{F1} & 
 No. CWEs &\textbf{HR} & \textbf{F1} \\
 \hline
 1 & 8 & 91.80 & 91.11 &  8 & 91.24  & 91.15  & 9 & 90.77 & 90.31 \\
 2 & 18 & 94.55 & 91.38 & 30 & 93.57 & 87.48  &  43 & 91.11 & 85.89  \\
 3 & 14 & 97.51 & 96.08 & 36 & 95.46  & 85.89 &  63 & 94.24 & 84.91 \\
 4 & 6 & 95.22 & 94.35 &  13 & 94.02 & 85.21 & 27& 94.57 & 90.81 \\
 5 & - & - & - & 3 & 91.53 & 91.53 &  7 & 90.80 & 90.96 \\
 
 \hline
 \end{tabular}
 \end{table*}

The hierarchical model offers the researchers a variety of levels of classification granularity. In other words, different levels of information may be required by researchers to characterize the properties of CVEs. As a result, providing level-specific knowledge about CVEs can offer various aspects of CVE characteristics. CVEs, on the other hand, may exploit more than one weakness, which cannot be addressed if they are mapped to only one CWE. As an outcome, hierarchical models are highly functional in recognizing such weaknesses and providing broader insight into exploitation to help in better assessment and, consequently, defense planning.
As previously noted, CVEs may be linked with multiple CWEs, however, NVD has only assigned them one. The similarity between CVE characteristics that are mapped to different CWEs and possibly the similarity across CWEs leads to different classification outcomes that are not necessarily incorrect. Since there is no ground truth for the additional corresponding CWEs with CVEs, we have conducted another experiment trying to show the performance of the model in classifying CVE to the correct class in "top K" model classification outputs. For this experiment, we used standard information retrieval metrics including hit rate at K (HR@K), precision at K (P@K), and recall at K (R@K).

Similar to CVSS prediction section, we carried out a comparison study to assess how well ChatGPT can predict the CWE associated with a given CVE description. For this purpose, we randomly chose 100 CVE samples disclosed recently and had ChatGPT classify them based on their corresponding CWE. 
Our examination demonstrated that ChatGPT achieved a $56\%$ accuracy in accurately classifying to the CWE class, while CVEDrill showcased a notably higher accuracy of $82\%$. Our analysis indicates that in various instances, ChatGPT struggles to comprehend the accurate impact and the weakness utilized to exploit a CVE, based on the provided description. Table \ref{tab: cgpt_cwe} in Appendix presents ten CVE instances illustrating cases where ChatGPT either successfully or unsuccessfully classified to the correct CWE.

\begin{table}[!ht]
\scriptsize
\centering
 \caption{SecureBERT performance evaluation on the hierarchical classification of CVEs to top 25, 50, and 100 CWEs.}
 \label{tab: cve2cwehir_eval}
\begin{tabular}{|l|c|c|c|c|c|c|c|}
 \hline
  \multicolumn{8}{|c|}{\textbf{Top 25 CWEs}}\\
 \hline
 L& No. CWEs &\textbf{HR@2} & \textbf{P@2} & \textbf{R@2} & 
 \textbf{HR@3} & \textbf{P@3} & \textbf{R@3} \\
 \hline
 1 & 8  & 96.82 & 94.44 & 97.44 & 98.92  & 95.83 & 94.29 \\
 2 & 18 & 94.95 & 94.13 & 90.81 & 95.11  & 94.54 & 90.79 \\
 3 & 14 & 97.51 & 96.63 & 95.69 & 97.51  & 96.69 & 95.63 \\
 4 & 6  & 95.42 & 95.63 & 93.18 & 95.61  & 95.12 & 95.06 \\
 \hline
  \end{tabular}
  \smallskip
 
 \begin{tabular}{|l|c|c|c|c|c|c|c|}
 \hline
 \multicolumn{8}{|c|}{\textbf{Top 50 CWEs}}\\
 \hline

 L& No. CWEs &\textbf{HR@2} & \textbf{P@2} & \textbf{R@2} & 
 \textbf{HR@3} & \textbf{P@3} & \textbf{R@3} \\
 \hline
 1 & 8  & 95.91 & 94.41 & 92.77 &  98.18 & 96.46 & 93.72\\
 2 & 30 & 94.29 & 89.84 & 86.78 &  94.33 & 91.55 & 87.58\\
 3 & 36 & 95.26 & 86.25 & 85.95 &  95.26 & 91.31 & 88.60\\
 4 & 13 & 94.73 & 90.44 & 88.62 &  94.73 & 92.02 & 86.27\\
 5 & 3  & 91.53 & 91.09 & 89.24 &  1.0   & 1.0 & 1.0\\
 \hline
 \end{tabular}
  \smallskip

 \begin{tabular}{|l|c|c|c|c|c|c|c|}
 \hline
  \multicolumn{8}{|c|}{\textbf{Top 100 CWEs}}\\
 \hline
 L& No. CWEs &\textbf{HR@2} & \textbf{P@2} & \textbf{R@2} & 
 \textbf{HR@3} & \textbf{P@3} & \textbf{R@3} \\
 \hline
 
 1 & 9  &  95.12 &   94.21  &  93.38 &  97.51 &  95.17  &     91.20 \\
 2 & 43 &  92.30 &   87.11  &  87.59 &  92.30 &  89.55  &     88.59 \\
 3 & 63 &  94.61 &   87.73  &  85.55 &  94.61 &  88.31  &     86.16 \\
 4 & 27 &  95.44 &   90.81  &  91.74 &  95.44 &  88.96  &     91.74 \\
 5 & 7  &  91.11 &   88.93  &  86.71 &  91.11 &  89.49  &     88.19 \\
 \hline
 \end{tabular}
  \smallskip
\end{table}

\subsection{CVE to Vulnerability Type (VT) Classification Model}
\begin{figure}
 \center
  \includegraphics[width=8cm]{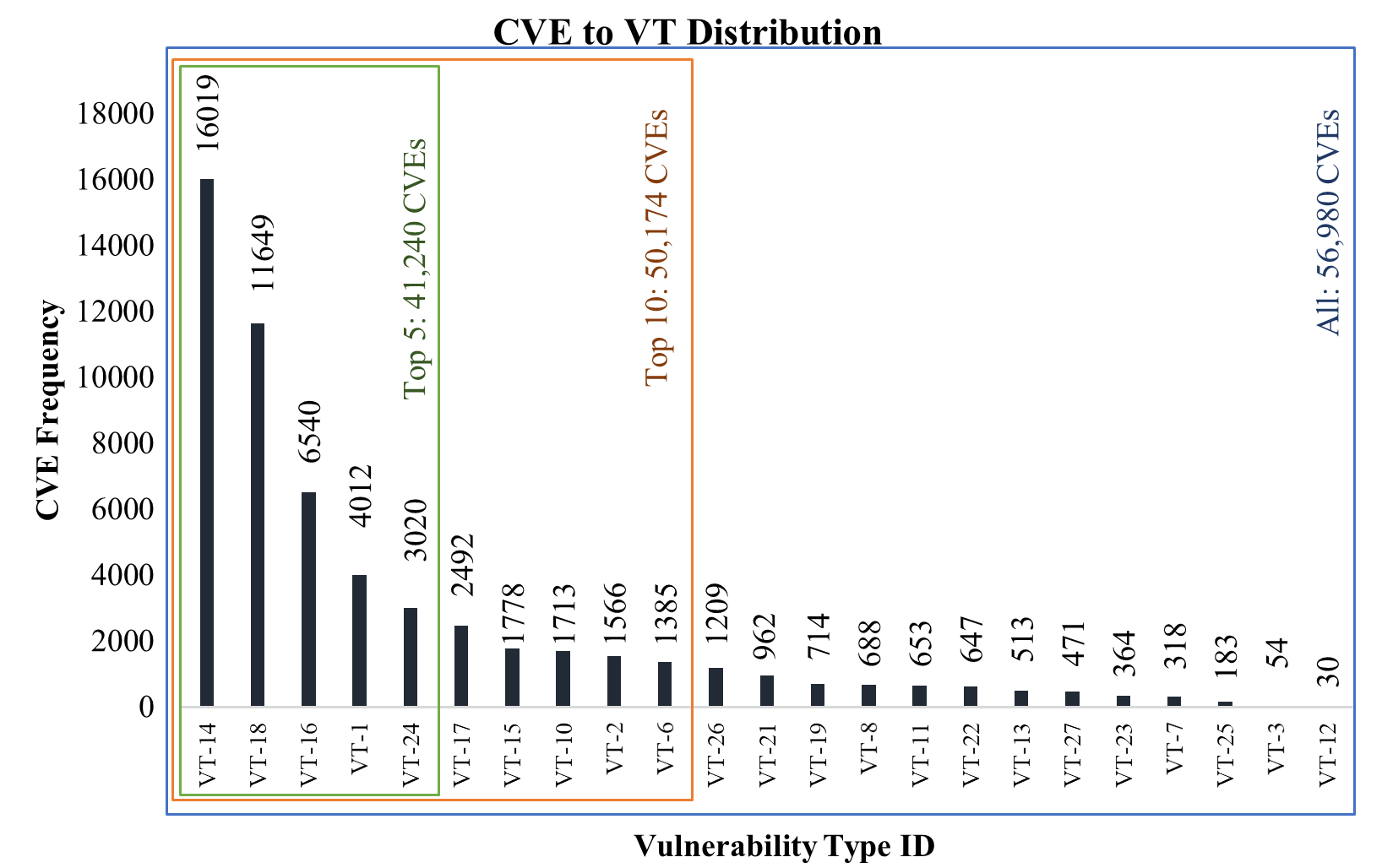}
  \caption{Distribution of CWEs in CVE to vulnerability type classification.}
  \label{fig: TopVT_dist}
\end{figure}

Similar to individual CWEs, VTs are  not distributed uniformly in NVD dataset. As depicted in Fig. \ref{fig: TopVT_dist}, about $72.3\%$ (41,240) and $88.03\%$ (50,174) of the CVEs are associated with top 5 and top 10 VTs, respectively. In addition, four VT IDs including 4, 5, 9, and 20 do not have any corresponding CVE in the NVD dataset. For training the CVE to VTs, we use $70\%$ of the available CVE descriptions as the training set to train the model to classify to one of the top 5, 10, and all 23 VTs provided in Table \ref{tab: cwe2vtype}.

\begin{table}[ht]
\centering
 \caption{SecureBERT performance evaluation in classifying CVEs to top 5, 10, and all vulnerability types.}
 \label{tab: cve2vt_eval}
\begin{tabular}{|l|c|c|c|c|c|c|c|}
 \hline
 Top VT & \textbf{HR@1} & \textbf{P@1} & \textbf{R@1} &
 \textbf{HR@2} & \textbf{P@2} & \textbf{R@2}\\
 \hline
 Top 5      &  93.12  &     91.34 &     89.08 &     95.21 &     92.31 &     89.46\\
  Top 10    &  91.66  &     88.63 &     86.79 &     93.71 &     90.03 &     87.11\\
   All      &  87.53  &     85.79 &     84.56 &     90.35 &     87.32 &     86.80\\
 \hline
 \end{tabular}
 \end{table}
Table \ref{tab: cve2vt_eval} shows the performance of our model. This table similarly reports the hit rate (HR), precision (P), and recall (R) in the top one and top two predictions of the proposed model. 
\section{Related Works}\label{sec:Related Works}
\subsubsection*{\textbf{Automated CVSS Prediction}}
To the best of our knowledge, there are only a few efforts in automating CVSS prediction in recent years. Khazaei \textit{et al}~\cite{khazaei2016automatic} introduced an objective method for CVSS V2 score calculation by extracting textual features from CVE descriptions and employing SVM, Random Forest, and fuzzy systems for prediction. In this study, the descriptions are tokenized, and after performing standard text mining preprocessing such as filtering the stop-words and stemming, the TF-IDF score of words is calculated, and the dimension is reduced using PCA and LDA techniques. The CVSS V2 scores are roughly predicted within the interval of $[i,i+1)$ where $i\in\{0,1,2,...,9\}$. This works reported $88.37\%$ accuracy through testing the fuzzy system implementation. This work leveraged an off-the-shelf machine learning model and failed to consider and report many important metrics in CVSS prediction. This work exclusively focused on finding the direct relationship between CVE description and the final score, while original scores are calculated based on different CVSS metric values where each metric specifies a variable in the non-linear score calculation formula. This direct method performs as a classification task which is problematic since one word (feature) might have a different impact on each metric, and training every metric in one shot is error-prone.
In the meantime, the bag-of-word approach does not take the context and semantics into account in to-be-announced CVE reports, such as synonym words and abbreviations.
In addition, there is no evidence showing the performance of this tool in predicting different value classes since different classes have a different number of samples, and predicting those who have less number of samples is important. 

In a similar approach, Elbaz \textit{et al}. \cite{elbaz2020fighting} implemented another technique based on linear regression to automatically predict the CVSS vector of newly disclosed vulnerabilities using their textual description. They used a similar bag-of-word approach and represented each CVE by word frequency vector. After filtering stop-words, two different dimension reduction approaches were applied to select the most predictive features from the corpus. In the first approach, the software vendor, product name, and software target exist in the Common Platform Enumeration (CPE), and all the words in CWE titles are collected, and a white list is generated to find and discard irrelevant features. In the second approach, the conditional entropy score for each word associated with each CVSS metric is calculated, and the top N words with the lowest score are selected for evaluation. Then a linear regression model is created to predict the scores for each CVSS metric. The accuracy of predicting each metric is varied from $60\%$ to $95\%$. This study suffers from certain weaknesses. 

First, similar to the previous work, the features are limited to the keywords extracted from CVE reports and CWE titles. This corpus fails to cover context and accordingly, semantic features such as synonyms, abbreviations, and similar words. Hence, there might be some less frequent words uniformly distributed among different values of one metric with a strong relationship with other words, which is a key signature for one specific value. The addressed feature extraction method in this study cannot find such a relationship, and therefore, it will assign a high entropy score to this word, and it will automatically be ignored in the classification. In addition, a limited corpus without semantic feature analysis may not handle a new word or different writing standard, which is very likely to have since different vendors have different writing styles and new concepts might be raised in the future CVE description.
In addition, the evaluation is incomplete and failed to address some important points. About half of the metrics are imbalanced, which means that the frequency of instances for one metric is much higher than the others. E.g, in \textit{Attack Vector}, there are $\sim35K$ instances with \textit{Network} value, while there are less than $500$ samples with \textit{Physical} value out of $28,000$ existing labeled CVEs. This can easily lead to poor predictive performance, specifically for the minority class during training, which has not been addressed in this study. 

\subsubsection*{\textbf{Automated CVE to CWE Classification}}
A great effort has been made initially by MITRE and NVD\footnote{National Vulnerability Database} to manually classify some CVEs to one or more CWE classes, each one shows critical shortcomings though.  Considering the growing number of CVEs and the high labor cost of manual classification, MITRE has classified 2553 CVEs (out of ~116K) to 364 CWE classes (out of 719) \cite{(3)CweMITRE2018}. On the other hand, NVD has \cite{(9)NVD2018} attempted to increase the quantity of CVE classification by mapping about 85,000 CVEs. Although MITRE has classified a smaller number of CVEs compared with NVD, it considers a higher number of CWEs and performs hierarchical and a more fine-grain classification. In the meantime, NVD classified more CVEs but it took a smaller number of CWEs into the account, without addressing the hierarchy. 

In the meantime, there have been several research efforts other than MITRE and NVD to analyze CVEs to enhance the searching process and to perform CVE categorization.
Aghaei \textit{et al.}~\cite{aghaei2020threatzoom, aghaei2019threatzoom} proposed a hierarchical classification model by utilizing the TF-IDF weights of N-grams extracted from CVE descriptions as the initial weights of a simple feed-forward neural network. This neural network is trained on both MITRE and NVD datasets in a hierarchical fashion and reported an accuracy between 75\% and 92\%. This work is tested on a limited and selected number of CVEs (10,000 CVEs) and failed to report the model performance at each level. In addition, vulnerability type classification has not been conducted in this work.

Neuhaus \textit{et al.}~\cite{(22)Neuhaus2010} proposed a semi-automatic method to analyze the CVE descriptions using topic models to find prevalent weaknesses and new trends. The test result reports 28 topics in these entries using Latent Dirichlet Allocation (LDA) and assigned LDA topics to CWEs~\cite{(22)Neuhaus2010}. This approach shows a highly limited accuracy depending on the CWE type.

Na \textit{et al.}~\cite{(23)Na2016} proposed  Na\"ive Bayes classifier to categorize CVE entries into the top ten most frequently used CWEs with an accuracy of 75.5\%. However, the accuracy of this limited classification significantly decreases as the number of the considered CWEs increases (i.e., accuracy decreased from 99.8\% to 75.5\% when the number of CWE classes increased from 2 to 10). In addition, this approach does not consider the hierarchical structure of CWEs, which significantly limits their value.
Another classifier was developed to estimate the vulnerabilities in CVEs using the basis of previously identified ones by Rahman \textit{et al.} \cite{(24)Rehman2012}. This approach uses the Term Frequency-Inverse Document Frequency (TF-IDF) to assign weights to text tokens from the feature vector and Support Vector Machine (SVM) to map CVEs to CWEs. However, they use only six CWE classes and 427 CVE instances. In addition, their classifier does not follow the hierarchical structure for CWEs as well. All these addressed issues and limitations are resolved in this work.

\section{Conclusion}
This research has brought about a significant advancement in cybersecurity automation with the introduction of CVEDrill, an innovative predictive model and tool for CVE assessment and classification. The conventional manual approach to CVE analysis and threat prioritization has long been plagued by inefficiencies and potential inaccuracies. Our methodology directly addresses these challenges by presenting a streamlined and automated solution that enhances the precision, speed, and comprehensiveness of CVE analysis.

By harnessing domain-specific language models, CVEDrill showcases a remarkable ability to predict CVSS vectors, enabling accurate prioritization and precise estimation of threat impact for proactive cybersecurity measures. Furthermore, its seamless integration of CWE classification empowers organizations to efficiently categorize vulnerabilities and identify potential mitigations, eliminating the bottlenecks associated with manual analysis.

The value of our work extends to various dimensions of cybersecurity operations. Organizations can now make well-informed decisions about service patching, security hardening, and other countermeasures, driven by data-driven insights rather than relying on labor-intensive and error-prone manual processes. 
In comparison to existing tools like ChatGPT, CVEDrill stands out as a comprehensive, accurate, and domain-specific solution, setting a new standard for CVE analysis and vulnerability management. Specifically, in extensive testing with over 100 arbitrary CVE descriptions, CVEDrill demonstrated an average accuracy of $83\%$, surpassing ChatGPT, which reported an average accuracy of $59\%$ in predicting CVSS vectors.
Moreover, CVEDrill achieved an accuracy rate of $82\%$ in accurately classifying the correct CWE class for the provided set of 100 randomly selected CVE samples. This performance surpasses that of ChatGPT, which achieved a lower accuracy rate of $56\%$ for the same classification task. 

This performance showcases the strength and reliability of CVEDrill in vulnerability assessment and analysis. 
As we progress, we expect that CVEDrill will play a pivotal role in enhancing proactive cybersecurity efforts, fortifying digital landscapes against emerging threats, and contributing to standardization and common practices.
\bibliographystyle{IEEEtranS.bst}
\bibliography{references.bib}
\begin{table*}[!ht]
\centering
\centerline{\Large APPENDIX}
\vspace{0.2in}
\small 
\begin{tabular}{|p{10cm}|*{8}{c|}c|}
\hline
\textbf{Random CVEs disclosed in 2023 (unseen CVEs in ChatGPT )} & \textbf{AV} & \textbf{AC} & \textbf{PR} & \textbf{UI} & \textbf{S} & \textbf{C} & \textbf{I} & \textbf{A} & \textbf{CWE} \\
\hline
Multiple command injection vulnerabilities could lead to unauthenticated remote code execution by sending specially crafted packets destined to the PAPI (Aruba Networks access point management protocol) UDP port (8211). Successful exploitation of these vulnerabilities results in the ability to execute arbitrary code as a privileged user on the underlying operating system. & {\color[HTML]{4B5320} \textbf{x}} & {\color[HTML]{4B5320} \textbf{x}} & {\color[HTML]{4B5320} \textbf{x}} & {\color[HTML]{4B5320} \textbf{x}} & {\color[HTML]{4B5320} \textbf{x}} &  & {\color[HTML]{4B5320} \textbf{x}} & {\color[HTML]{4B5320} \textbf{x}} & {\color[HTML]{960018} \textbf{0.87}} \\
\hline
Sensitive data could be exposed in logs of cloud-init before version 23.1.2. An attacker could use this information to find hashed passwords and possibly escalate their privilege & {\color[HTML]{4B5320} \textbf{x}} &  & {\color[HTML]{4B5320} \textbf{x}} & {\color[HTML]{4B5320} \textbf{x}} &  & {\color[HTML]{4B5320} \textbf{x}} & {\color[HTML]{4B5320} \textbf{x}} & {\color[HTML]{4B5320} \textbf{x}} & {\color[HTML]{960018} \textbf{0.75}} \\
\hline
IBM Sterling B2B Integrator Standard Edition 6.0.0.0 through 6.1.2.1 is vulnerable to SQL injection. A remote attacker could send specially crafted SQL statements, which could allow the attacker to view, add, modify or delete information in the back-end database. & {\color[HTML]{4B5320} \textbf{x}} & {\color[HTML]{4B5320} \textbf{x}} &  & {\color[HTML]{4B5320} \textbf{x}} &  & {\color[HTML]{4B5320} \textbf{x}} & {\color[HTML]{4B5320} \textbf{x}} & & {\color[HTML]{960018} \textbf{0.63}}\\
\hline
Improper usage of implicit intent in Bluetooth prior to SMR Mar-2023 Release 1 allows attacker to get MAC address of connected device. &  & {\color[HTML]{4B5320} \textbf{x}} &  & {\color[HTML]{4B5320} \textbf{x}} & {\color[HTML]{4B5320} \textbf{x}} & {\color[HTML]{4B5320} \textbf{x}} & {\color[HTML]{4B5320} \textbf{x}} & {\color[HTML]{4B5320} \textbf{x}} & {\color[HTML]{960018} \textbf{0.75}}\\
\hline
*Pimcore is an open source data and experience management platform. Prior to version 10.5.19, quoting is not done properly in UUID DAO model. There is the theoretical possibility to inject custom SQL if the developer is using this method with input data and not doing proper input validation in advance and so relies on the auto-quoting being done by the DAO class. Users should update to version 10.5.19 to receive a patch or, as a workaround, apply the patch manually. &  & {\color[HTML]{4B5320} \textbf{x}} & {\color[HTML]{4B5320} \textbf{x}} & {\color[HTML]{4B5320} \textbf{x}} & {\color[HTML]{4B5320} \textbf{x}} &  &  & & {\color[HTML]{960018} \textbf{0.5}}\\
\hline
Flatpak is a system for building, distributing, and running sandboxed desktop applications on Linux. Versions prior to 1.10.8, 1.12.8, 1.14.4, and 1.15.4 contain a vulnerability similar to CVE-2017-5226, but using the `TIOCLINUX` ioctl command instead of `TIOCSTI`. If a Flatpak app is run on a Linux virtual console such as `/dev/tty1`, it can copy text from the virtual console and paste it into the command buffer, from which the command might be run after the Flatpak app has exited. Ordinary graphical terminal emulators like xterm, gnome-terminal and Konsole are unaffected. This vulnerability is specific to the Linux virtual consoles `/dev/tty1`, `/dev/tty2` and so on. A patch is available in versions 1.10.8, 1.12.8, 1.14.4, and 1.15.4. As a workaround, don't run Flatpak on a Linux virtual console. Flatpak is primarily designed to be used in a Wayland or X11 graphical environment. & {\color[HTML]{4B5320} \textbf{x}} & {\color[HTML]{4B5320} \textbf{x}} & {\color[HTML]{4B5320} \textbf{x}} & {\color[HTML]{4B5320} \textbf{x}} &  &  & {\color[HTML]{4B5320} \textbf{x}} & & {\color[HTML]{960018} \textbf{0.63}}\\
\hline
*In onPackageAddedInternal of PermissionManagerService.java, there is a possible way to silently grant a permission after a Target SDK update due to a permissions bypass. This could lead to local escalation of privilege after updating an app to a higher Target SDK with no additional execution privileges needed. User interaction is not needed for exploitation. Product: AndroidVersions: Android-11 Android-12 Android-12L Android-13Android ID: A-221040577 &  & {\color[HTML]{4B5320} \textbf{x}} & {\color[HTML]{4B5320} \textbf{x}} & {\color[HTML]{4B5320} \textbf{x}} & {\color[HTML]{4B5320} \textbf{x}} &  &  & & {\color[HTML]{960018} \textbf{0.5}}\\
\hline
A buffer overflow vulnerability was found in the Netfilter subsystem in the Linux Kernel. This issue could allow the leakage of both stack and heap addresses, and potentially allow Local Privilege Escalation to the root user via arbitrary code execution. & {\color[HTML]{4B5320} \textbf{x}} & {\color[HTML]{4B5320} \textbf{x}} & {\color[HTML]{4B5320} \textbf{x}} & {\color[HTML]{4B5320} \textbf{x}} & {\color[HTML]{4B5320} \textbf{x}} &  & {\color[HTML]{4B5320} \textbf{x}} & & {\color[HTML]{960018} \textbf{0.75}}\\
\hline
*WordPress through 6.1.1 depends on unpredictable client visits to cause wp-cron.php execution and the resulting security updates, and the source code describes "the scenario where a site may not receive enough visits to execute scheduled tasks in a timely manner," but neither the installation guide nor the security guide mentions this default behavior, or alerts the user about security risks on installations with very few visits. &  &  & {\color[HTML]{4B5320} \textbf{x}} & {\color[HTML]{4B5320} \textbf{x}} & {\color[HTML]{4B5320} \textbf{x}} &  & {\color[HTML]{4B5320} \textbf{x}} & & {\color[HTML]{960018} \textbf{0.5}}\\
\hline
Microsoft SharePoint Server Remote Code Execution Vulnerability & {\color[HTML]{4B5320} } & {\color[HTML]{4B5320} \textbf{x}} & {\color[HTML]{4B5320} \textbf{x}} & {\color[HTML]{4B5320} \textbf{x}} & {\color[HTML]{4B5320} } &  & {\color[HTML]{4B5320} \textbf{x}} & {\color[HTML]{4B5320} \textbf{x}} & {\color[HTML]{960018} \textbf{0.75}}\\
\hline
\end{tabular}
\caption{Examples of CVSS prediction using ChatGPT.}
\label{Tab: chatgpt_cvss}
\end{table*}

\begin{table*}[]
\centering
\vspace{0.2in}
\small 
\begin{tabular}{|p{12cm}|l|l|}
\hline
\textbf{CVE} & \textbf{Predicted}             & \textbf{Correct} \\ \hline
CVE-2023-22747: There are multiple command injection vulnerabilities that could lead to unauthenticated remote code execution by sending specially crafted packets destined to the PAPI (Aruba Networks access point management protocol) UDP port (8211). Successful exploitation of these vulnerabilities result in the ability to execute arbitrary code as a privileged user on the underlying operating system.                                                                                                                                                                                                                                                                                                  & CWE-78                         & CWE-77           \\ \hline
CVE-2023-1786: Sensitive data could be exposed in logs of cloud-init before version 23.1.2. An attacker could use this information to find hashed passwords and possibly escalate their privilege.                                                                                                                                                                                                                                                                                                                                                                                                                                                                                                                    & {\color[HTML]{036400} CWE-532} & CWE-532          \\ \hline
CVE-2022-22338: IBM Sterling B2B Integrator Standard Edition 6.0.0.0 through 6.1.2.1 is vulnerable to SQL injection. A remote attacker could send specially crafted SQL statements, which could allow the attacker to view, add, modify or delete information in the back-end database. IBM X-Force ID: 219510.                                                                                                                                                                                                                                                                                                                                                                                                       & {\color[HTML]{036400} CWE-89}  & CWE-89           \\ \hline
CVE-2023-21452: Improper usage of implicit intent in Bluetooth prior to SMR Mar-2023 Release 1 allows attacker to get MAC address of connected device.                                                                                                                                                                                                                                                                                                                                                                                                                                                                                                                                                                & CWE-284                        & CWE-285          \\ \hline
CVE-2023-28108: Pimcore is an open source data and experience management platform. Prior to version 10.5.19, quoting is not done properly in UUID DAO model. There is the theoretical possibility to inject custom SQL if the developer is using this methods with input data and not doing proper input validation in advance and so relies on the auto-quoting being done by the DAO class. Users should update to version 10.5.19 to receive a patch or, as a workaround, apply the patch manually.                                                                                                                                                                                                                & {\color[HTML]{036400} CWE-89}  & CWE-89           \\ \hline
CVE-2023-28100: Flatpak is a system for building, distributing, and running sandboxed desktop applications on Linux. Versions prior to 1.10.8, 1.12.8, 1.14.4, and 1.15.4 contain a vulnerability similar to CVE-2017-5226, but using the `TIOCLINUX` ioctl command instead of `TIOCSTI`. If a Flatpak app is run on a Linux virtual console such as `/dev/tty1`, it can copy text from the virtual console and paste it into the command buffer, from which the command might be run after the Flatpak app has exited. Ordinary graphical terminal emulators like xterm, gnome-terminal and Konsole are unaffected. This vulnerability is specific to the Linux virtual consoles `/dev/tty1`, `/dev/tty2` and so on. & CWE-434                        & CWE-20           \\ \hline
CVE-2022-40679: An improper neutralization of special elements used in an OS command vulnerability {[}CWE-78{]} in FortiADC 5.x all versions, 6.0 all versions, 6.1 all versions, 6.2.0 through 6.2.4, 7.0.0 through 7.0.3, 7.1.0; FortiDDoS 4.x all versions, 5.0 all versions, 5.1 all versions, 5.2 all versions, 5.3 all versions, 5.4 all versions, 5.5 all versions, 5.6 all versions and FortiDDoS-F 6.4.0, 6.3.0 through 6.3.3, 6.2.0 through 6.2.2, 6.1.0 through 6.1.4 may allow an authenticated attacker to execute unauthorized commands via specifically crafted arguments to existing commands.                                                                                                        & {\color[HTML]{036400} CWE-78}  & CWE-78           \\ \hline
CVE-2023-0179: A buffer overflow vulnerability was found in the Netfilter subsystem in the Linux Kernel. This issue could allow the leakage of both stack and heap addresses, and potentially allow Local Privilege Escalation to the root user via arbitrary code execution.                                                                                                                                                                                                                                                                                                                                                                                                                                         & CWE-122                        & CWE-190          \\ \hline
CVE-2023-20102: A vulnerability in the web-based management interface of Cisco Secure Network Analytics could allow an authenticated, remote attacker to execute arbitrary code on the underlying operating system. This vulnerability is due to insufficient sanitization of user-provided data that is parsed into system memory. An attacker could exploit this vulnerability by sending a crafted HTTP request to an affected device. A successful exploit could allow the attacker to execute arbitrary code on the underlying operating system as the administrator user.                                                                                                                                       & CWE-20                         & CWE-502          \\ \hline
CVE-2023-0179: A buffer overflow vulnerability was found in the Netfilter subsystem in the Linux Kernel. This issue could allow the leakage of both stack and heap addresses, and potentially allow Local Privilege Escalation to the root user via arbitrary code execution.                                                                                                                                                                                                                                                                                                                                                                                                                                         & CWE-787                        & CWE-190          \\ \hline
\end{tabular}

\caption{Examples of CVE to CWE classification using ChatGPT.}
\label{tab: cgpt_cwe}
\end{table*}

\end{document}